\documentclass[reprint,nofootinbib,preprintnumbers,aps,prd,superscriptaddress,floatfix,amsmath,amssymb]{revtex4-2}

\usepackage{graphicx}
\usepackage{dcolumn}
\usepackage{bm}
\usepackage{braket}
\usepackage{breqn}
\usepackage{color}
\usepackage[legacycolonsymbols]{mathtools}

\usepackage[colorlinks=true,linkcolor=red,citecolor=blue]{hyperref}

\renewcommand{\Im}[1]{\mathrm{Im}\left[ #1 \right]}

\begin{document}

\preprint{RUP-24-12}

\title{
Self-interacting axion clouds around rotating black holes in binary systems}



\author{Takuya Takahashi}
\email{ttakahashi1@rikkyo.ac.jp}
\affiliation{Department of Physics$,$ Rikkyo University$,$ Toshima$,$ Tokyo 171-8501$,$ Japan}
\author{Hidetoshi Omiya}
\email{omiya@tap.scphys.kyoto-u.ac.jp}
\affiliation{Department of Physics$,$ Kyoto University$,$ Kyoto 606-8502$,$ Japan}
\author{Takahiro Tanaka}
\email{t.tanaka@tap.scphys.kyoto-u.ac.jp}
\affiliation{Department of Physics$,$ Kyoto University$,$ Kyoto 606-8502$,$ Japan}
\affiliation{Center for Gravitational Physics and Qunatum Information$,$ Yukawa Institute for Theoretical Physics$,$ Kyoto University$,$ Kyoto 606-8502$,$ Japan}


\date{\today}

\begin{abstract}
Superradiant instability can form clouds around rotating black holes (BHs) composed of ultralight bosonic fields, such as axions. A BH with such a cloud in a binary system exhibits rich phenomena, and gravitational waves (GWs) from the BH merger provide a means to probe axions. For the first time, we study the evolution of axion clouds in a binary system during the inspiral phase, including axion self-interaction effects. When the self-interaction is significant, unlike in the negligible case, two types of clouds coexist through mode coupling. We examine the evolution of the system considering the effects of dissipation caused by both self-interaction and tidal interaction. For tidal interaction, in addition to the processes of emission to infinity and absorption by the BH, indirect emission via transitions (both resonant and off-resonant) is also considered as a second-order perturbation. Our results demonstrate that the signatures of axion self-interaction are imprinted in the modification of the GW phase. Furthermore, we find the possibility of a dynamical instability called bosenova during the binary inspiral phase.  
\end{abstract}


\maketitle

\section{Introduction}\label{Sec:intro}
The development of gravitational wave (GW) astronomy provides a novel means to explore the light sector of particle physics.
In particular, axions or axion-like particles are reasonably proposed as an extension of the Standard Model with very light masses.
They were originally introduced to solve the strong CP problem~\cite{Peccei:1977hh, Peccei:1977hh, Weinberg:1977ma, Wilczek:1977pj, Kim:1979if} and are considered viable candidates for dark matter~\cite{Dine:1982ah, Preskill:1982cy, Abbott:1982af, Hui:2016ltb}.
In addition, string theory predicts the existence of numerous axions in the so-called string axiverse scenario~\cite{Arvanitaki:2009fg, Svrcek:2006yi, Mehta:2021pwf}. 

A bosonic field with a mass around a rotating black hole (BH) can trigger the superradiant instability that extracts energy and angular momentum from the BH~\cite{Press:1972zz, Brito:2020oca}.
This effect is most pronounced when the Compton wavelength of the field is comparable to the BH size.
For astrophysical BHs, this corresponds to the mass ranging from $10^{-20}$eV to $10^{-10}$eV.
Bosonic field bound by the BH's gravity grows unstably through superradiance, leading to the formation of a macroscopic condensate~\cite{Arvanitaki:2010sy, Zouros:1979iw, Detweiler:1980uk, Dolan:2007mj}.
We consider axions as such bosons and refer to this condensate as an ``axion cloud''.
The presence of the cloud shows observational signatures, including the BH spin-down~\cite{Arvanitaki:2010sy, Brito:2014wla, Stott:2020gjj}, impacts on the shadow~\cite{Roy:2021uye,Chen:2019fsq, Chen:2021lvo, Chen:2022nbb, Chen:2022oad} and the emission of continuous GWs~\cite{Arvanitaki:2014wva, Arvanitaki:2016qwi, Yoshino:2013ofa, Brito:2017zvb, Isi:2018pzk, Ng:2020jqd, Siemonsen:2022yyf, Collaviti:2024mvh, Omiya:2024xlz}.

Signature in GWs from a binary coalescence is one of the most interesting subjects in exploring axion clouds.
In a binary system, the axion cloud is affected by the tidal interaction from the companion.
As a radiation reaction of GW emission from the orbital motion, the separation of the binary decreases, and the orbital frequency increases gradually.
Resonant transitions occur when the orbital frequency coincides with the difference between eigenfrequencies of the axion~\cite{Baumann:2018vus, Baumann:2019ztm}.
There is also absorption by the BH due to transitions to modes with high decay rates~\cite{Tong:2022bbl, Takahashi:2023flk}.
Furthermore, if the orbital frequency is sufficiently large, axions can be excited to unbound states and emitted to infinity~\cite{Takahashi:2021yhy, Baumann:2021fkf, Baumann:2022pkl}.

Simultaneously, the orbital evolution is altered by the transport of angular momentum from the axion cloud. 
This signal, indicating the presence of the cloud, appears in the phase evolution of GWs and has been extensively studied~\cite{Baryakhtar:2022hbu}.
For instance, the evolution of the binary system with a generic orbit has been investigated, providing characteristic insights into orbital eccentricity and inclination angle~\cite{Tomaselli:2023ysb, Tomaselli:2024bdd, Tomaselli:2024dbw, Fan:2023jjj, Boskovic:2024fga, Kim:2024rgf}.
A relativistic formulation based on BH perturbation theory has also been established~\cite{Brito:2023pyl, Cannizzaro:2023jle, Duque:2023cac, DeLuca:2021ite, DeLuca:2022xlz}.
Furthermore, the signatures that appear by considering astrophysical scenarios have also been discussed~\cite{Hui:2022sri, Du:2022trq, Guo:2023lbv, Guo:2024iye}.

Generally, axions interact with other particles, including self-interaction. (See, {\it e.g.}, Refs.~\cite{Sakurai:2023hkg, Spieksma:2023vwl, Sarmah:2024nst, Chen:2023vkq} for the coupling to the photon or neutrinos and Refs.~\cite{Chakrabarti:2022owq, Dave:2023wjq, Dave:2023egr, Kadota:2023wlm} for the self-interaction in the case where the axion is dark matter.)
However, the effects of self-interaction have not been considered in the context of binary systems so far.
In this paper, we investigate for the first time the evolution of the axion clouds in binary systems including the effects of self-interaction.
Depending on the strength of the self-interaction, this effect can become dominant over superradiance.
Consequently, as described below, novel signatures and phenomena can arise in binary systems.

In the absence of the self-interaction, the axion cloud tends to consist of the single fastest growing mode.
The cloud is thought to grow, extracting the BH spin, until the superradiance condition is saturated.
On the other hand, if the self-interaction is strong enough, another mode can be excited by energy transfer from the fastest growing mode due to a mode coupling~\cite{Gruzinov:2016hcq, Baryakhtar:2020gao, Omiya:2022gwu}. 
In balance with energy flow, the two clouds settle into a quasi-stationary state long before the BH spin is completely extracted.
In this case, the lifetime of the BH spin becomes longer, and the tidal interaction may become effective while the constituent BH is in the quasi-stationary state with the two clouds.
Thus, examining the evolution of two clouds with the self-interaction is necessary.

The effects of the tidal interaction on each cloud are quantitatively different.
Not only that, but the secondary cloud (specified by $l=m=2$) is also easily coupled with modes that fall into BH, leading to significant dissipation channels.
We calculate the energy and angular momentum fluxes induced by the self-interaction and the tidal interaction for each cloud in a perturbative manner.
Also, dissipation due to indirect emission to infinity from the intermediate bound state excited by the tidal perturbation is calculated using the second-order perturbation.
Then, we present the formulation that provides us with the adiabatic time evolution based on flux-balance laws.

Unlike the non-interacting case, the masses of the clouds in the quasi-equilibrium state depend on the decay constant, which characterizes the strength of the self-interaction.
Therefore, information on the decay constant is imprinted in the GW signal from the cloud as a modification to the phase evolution.
By solving the evolution equations, we show when the clouds are disrupted and how the orbital motion is affected.

In addition, the self-interaction can lead to a
more striking explosive phenomenon known as ``bosenova''.
This occurs due to the attractive nature of the self-interaction 
when the amplitude of the cloud exceeds a certain threshold~\cite{Arvanitaki:2010sy, Yoshino:2012kn, Yoshino:2015nsa}.
While the growth of the cloud saturates in general before reaching the threshold for isolated BH cases~\cite{Baryakhtar:2020gao, Omiya:2022gwu}, the bosenova may occur in binary systems.
This is because the secondary cloud is more easily affected by the tidal interaction, leading to earlier disruption than the primary cloud, which suppresses energy transfer between two clouds. 
We discuss the possibility of the bosenova occurrence during the inspiral phase of the binary system.

This paper is organized as follows.
In Sec.~\ref{Sec:review}, we review the basics of axion clouds around BHs formed by the superradiant instability.
In Sec.~\ref{sec:interaction}, we consider the effects of the self-interaction and the tidal interaction and show the fluxes induced by them.
In Sec.~\ref{sec:evolution_eq}, we give the adiabatic evolution equations of the system.
In Sec.~\ref{sec:results}, we show several consequences obtained from our calculation.
Finally, we give a conclusion in Sec.~\ref{sec:conslucion}.
Throughout this paper, we use the unit with $c=G=\hbar=1$, unless otherwise stated.

\section{Axion Clouds}\label{Sec:review}
In this section, we briefly review the basics of axion clouds around BHs.
We first consider a non-interacting real scalar field $\phi$ with mass $\mu$ around a Kerr BH with mass $M$ and the angular momentum $J=a M=\chi M^2$.
The equation of motion for a scalar field is given by
\begin{align}\label{eq:KG}
	(g^{\mu\nu}\nabla_{\mu}\nabla_{\nu}-\mu^2)\phi=0~,
\end{align}
where $g_{\mu\nu}$ is the Kerr metric.
In the non-relativistic regime, it is appropriate to adopt the ansatz
\begin{align}
	\phi=\frac{1}{\sqrt{2\mu}}\left(e^{-i\mu t}\psi+e^{i\mu t}\psi^{\ast}\right)~,
\end{align}
where $\psi$ is a complex scalar field whose timescale of variation is much longer than $\mu^{-1}$.
Substituting it into Eq.~\eqref{eq:KG}, the equation of motion reduces to
\begin{align}
	i\frac{\partial}{\partial t}\psi=\left(-\frac{1}{2\mu}\nabla^2-\frac{\alpha}{r}+{\cal O}(\alpha^2)\right)\psi~.
\end{align}
Here, we have introduced the gravitational fine structure constant
\begin{align}
	\alpha\coloneqq M\mu\simeq0.2\left(\frac{M}{30M_{\odot}}\right)\left(\frac{\mu}{10^{-12}{\rm eV}}\right)~.
\end{align}
Now, we are interested in the quasi-bound states that satisfy the ingoing boundary condition at the horizon and decay at infinity.
In the non-relativistic approximation, the solutions in the leading order are given by
\begin{align}
	\psi_{nlm}\propto e^{-i(\omega_{nlm}-\mu)t}R_{nl}(r)Y_{lm}(\theta,\varphi)~,
\end{align}
where $R_{nl}(r)$ are equivalent to the wavefunctions of a hydrogen atom, and $Y_{lm}(\theta,\varphi)$ are the spherical harmonics.
The integers, $n,l,m$, are the principal, angular momentum, and magnetic quantum numbers, respectively.
Here, we normalized the wavefunction and the spherical harmonics as $\int dr\, r^2 R_{nl}R_{n'l'}=\delta_{nn'}\delta_{ll'}$ and $\int d\cos\theta\, d\varphi\, Y_{lm}^\ast Y_{l'm'}=\delta_{ll'}\delta_{mm'}$.
Also, we define the particle number occupying the $\ket{nlm}$ state as $N_{nlm}=\int d^3\! x|\psi_{nlm}|^2$.
Eigenfrequencies are given by~\cite{Baumann:2019eav,Baumann:2018vus,Detweiler:1980uk,Pani:2012bp,Bao:2022hew} 
\begin{equation}
\omega_{nlm}=(\omega_{R})_{nlm}+i(\omega_{I})_{nlm}~,
\end{equation}
with
\begin{align}
(\omega_{R})_{nlm}&=\mu\left(1-\frac{\alpha^2}{2n^2}-\frac{\alpha^4}{8n^4}+\frac{(2l-3n+1)\alpha^4}{n^4(l+1/2)} \right. \notag \\
& \qquad \qquad \qquad \quad \left.+\frac{2m\chi\alpha^5}{n^3 l(l+1/2)(l+1)}\right) ~, \label{omegaR} \\
(\omega_{I})_{nlm}&=2\frac{r_{+}}{M}C_{nlm}(m\Omega_{H}-(\omega_R)_{nlm})\alpha^{4l+5}~, \label{omegaI}
\end{align}
where $r_{+}=M+\sqrt{M^2-a^2}$ is the horizon radius, 
$\Omega_{H}=a/2Mr_{+}$ is the angular velocity of the BH horizon 
and
\begin{align}
	C_{nlm}=&\frac{2^{4l+1}(n+l)!}{n^{2l+4}(n-l-1)!}\left(\frac{l!}{(2l)!(2l+1)!}\right)^2 \notag \\
	&\times\prod_{j=1}^l\left[j^2(1-\chi^2)+(\chi m-2r_+(\omega_R)_{nlm})^2\right]~.
\end{align}
Note that this approximation is well justified for $\alpha/l\ll1$.

From Eq.~\eqref{omegaI}, when $\omega_R<m\Omega_H$, the imaginary part of the eigenfrequency is positive and the amplitude of such a mode grows by the superradiance.
The critical value of the BH spin at which the superradiance condition is marginally satisfied is
\begin{align}
    \chi_{\rm crit}=\frac{4m(M\omega_R)}{m^2+4(M\omega_R)^2}~.
\end{align}
The growth rate is suppressed for large $m$ by the factor $\alpha^{4l}$ with $l\geq |m|$, while 
the critical value of the spin becomes smaller as $m$ increases.
The cloud that is composed of the fastest-growing mode is formed first.

\section{Self-interacting axion clouds in binary systems}\label{sec:interaction}
In this paper, we consider the evolution of the axion field with self-interaction around a BH in a binary system.
Thanks to the adiabatic growth of the cloud, we can treat the effects of these interactions in terms of energy and angular-momentum conservation laws.
Here, we summarize the dissipation processes involved in this system.

\subsection{Setup}
The action for an axion field around a BH in a binary system is
\begin{align}
	S=\int d^4x\sqrt{-\tilde{g}}\left(-\frac{1}{2}\tilde{g}^{\mu\nu}\partial_\mu\phi\partial_\nu\phi-V(\phi)\right)~.
\end{align}
The metric of the spacetime is given by
\begin{align}\tilde{g}_{\mu\nu}=g_{\mu\nu}+h_{\mu\nu}^{\rm tidal}~,
\end{align}
where $h_{\mu\nu}^{\rm tidal}$ is the tidal perturbation from the companion orbiting around the BH.
We adopt the potential given by
\begin{align}\label{eq:cos}
	V(\phi)&=\mu^2F_a^2\left(1-\cos\frac{\phi}{F_a}\right)\cr
 &\sim \frac{1}{2} \mu^2 \phi^2 - \frac{1}{4!}\mu^2 F_a^2\left(\frac{\phi}{F_a}\right)^4\,,
\end{align}
where $F_a$ is the decay constant.
For the self-interaction, as shown in Refs.~\cite{Omiya:2022mwv, Omiya:2022gwu}, the dominant contribution comes from the $\phi^4$ term in the expansion of Eq.~\eqref{eq:cos}.
Therefore, we truncate the self-interaction at the $\phi^4$ term.
In the non-relativistic regime, as in the previous section, the equation of motion reduces to
\begin{align}\label{eq:eom}
	\left(i\frac{\partial}{\partial t}+\frac{1}{2\mu}\nabla^2+\frac{\alpha}{r}\right)\psi=-\frac{1}{8F_a^2}|\psi|^2\psi+V_{\ast}\psi~,
\end{align}
where $V_{\ast}=\frac{1}{2}\mu h^{tt}_{\rm tidal}$.
Here, we ignored the rapidly oscillating term containing the factor $e^{\pm 2i\mu t}$, 
since their contribution is subdominant to the averaged dynamics.
We also ignore the GW emission from the cloud since its effect is much smaller than the self-interaction and the tidal effects.

In this paper, we focus on the case in which the $\ket{nlm}=\ket{211}$ mode is the fastest-growing mode.
Denoting the particle number of axions at this mode as
\begin{align}
	N_1=\int d^3x|\psi_{211}|^2~,
\end{align}
we investigate its evolution.
The flux induced by each term of the right-hand side of Eq.~\eqref{eq:eom} can be calculated perturbatively.
In App.~\ref{app:flux}, we summarize the formalism to calculate the flux for a general source ${\cal S}=e^{-i\omega_s t}S(\bm{x})$.
The results for the self-interaction and the tidal interaction are given in the following subsections.

\subsection{Self-interaction}

The first term of the right-hand side of Eq.~\eqref{eq:eom}, the self-interaction, induces the coupling with other modes.
The dominant process is involved in the coupling between the $\ket{211}$ and $\ket{322}$ modes.
Due to the non-linearity of the self-interaction, it is also necessary to solve the equation for
\begin{align}
N_2=\int d^3x\, |\psi_{322}|^2~.
\end{align}
The evolution of these modes is almost determined by the two processes: (1) dissipation to the BH associated with the transition to a non-superradiant $l=m=0$ mode, and (2) dissipation to infinity associated with the transition to an unbound mode with $l=m=3$ and $\omega>\mu$.
The corresponding number flux can be obtained as ${\cal F}_{\cal I}$ in Eq.~\eqref{eq:FI2} with the source term ${\cal S}=-\frac{1}{8F_a^2}\psi_{211}^{\ast}\psi_{322}^2$, and ${\cal F}_{\cal H}$ in Eq.~\eqref{eq:FH2} with the source term ${\cal S}=-\frac{1}{8F_a^2}\psi_{211}^2\psi_{322}^\ast$.
We include only these processes as the effects of self-interaction.
For other minor contributions such as relativistic emission and GW emission, see Ref.~\cite{Baryakhtar:2020gao} and~\cite{Omiya:2022gwu}.

Here, we denote the respective number fluxes as
\begin{align}
    {\cal F}_{\cal I}^{\rm SI}=f_3 N_1N_2^2~, \\
    {\cal F}_{\cal H}^{\rm SI}=f_0 N_1^2N_2~.
\end{align}
In the non-relativistic approximation, coefficients are given by~\cite{Baryakhtar:2020gao} 
\begin{align}
    f_0&=4.2\times 10^{-7}\alpha^7 \left(1+\sqrt{1-\chi^2}\right)\left(\frac{\mu}{F_a}\right)^4\mu~, \label{eq:f0nr} \\
    f_3&=1.1\times 10^{-8}\alpha^4 \left(\frac{\mu}{F_a}\right)^4\mu~.
\end{align}
The total changes in particle number in the $\ket{211}$ and $\ket{322}$ modes regarding the self-interaction are expressed as
\begin{align} \label{eq:ecSI}
    \dot{N}_1^{\rm SI}+\dot{N}_2^{\rm SI}&=-{\cal F}_{\cal I}^{\rm SI}-{\cal F}_{\cal H}^{\rm SI}~.
\end{align}
Also, the angular momentum flux is given by $(m/\omega_R){\cal F}^{\rm SI}$, where $\omega_R$ and $m$ are the frequency and magnetic quantum number of the dissipative mode, respectively. Thus, approximating $\omega_R\simeq\mu$, the change in angular momentum can be expressed as
\begin{align} \label{eq:acSI}
    \dot{N}_1^{\rm SI}+2\dot{N}_2^{\rm SI}&=-3{\cal F}_{\cal I}^{\rm SI}~.
\end{align} 
Defining the dissipative flux for each mode as ${\cal F}_i^{\rm SI}:=-\dot{N}_i^{\rm SI}$, we obtain
\begin{align} 
    {\cal F}_1^{\rm SI}&=2f_0N_1^2N_2-f_3N_1N_2^2~, \label{eq:Fself1} \\
    {\cal F}_2^{\rm SI}&=-f_0N_1^2N_2+2f_3N_2N_2^2~.\label{eq:Fself2}
\end{align}
from Eqs.~\eqref{eq:ecSI} and~\eqref{eq:acSI}.

For convenience, if we rewrite the energy flux divided by $M$ expressed in terms of the normalized cloud mass defined by $\tilde{M}_{{\rm c},i}\coloneqq M_{{\rm c},i}/M\coloneqq\mu N_i/M$, {\it i.e.,} $F_i^{\rm SI}=-\dot{\tilde{M}}_i$, we get
\begin{align}
    F_1^{\rm SI}&=2F_0\tilde{M}_1^2\tilde{M}_2-F_3\tilde{M}_1\tilde{M}_2^2~, \\
    F_2^{\rm SI}&=-F_0\tilde{M}_1^2\tilde{M}_2+2F_3\tilde{M}_1\tilde{M}_2^2~,
\end{align}
where $F_i\coloneqq(M/\mu)^2 f_i$, which are more explicitly written down as 
\begin{align}
    F_0&=4.2\times 10^{-7}\alpha^9 \left(1+\sqrt{1-\chi^2}\right)\left(\frac{M_{\rm Pl}}{F_a}\right)^4\mu~, \label{eq:F0nr} \\
    F_3&=1.1\times 10^{-8}\alpha^6 \left(\frac{M_{\rm Pl}}{F_a}\right)^4\mu~.
\end{align}
We tentatively recovered the Planck mass $M_{\rm Pl}=G^{-1/2}$ here.

\begin{figure}
\includegraphics[scale=0.8]{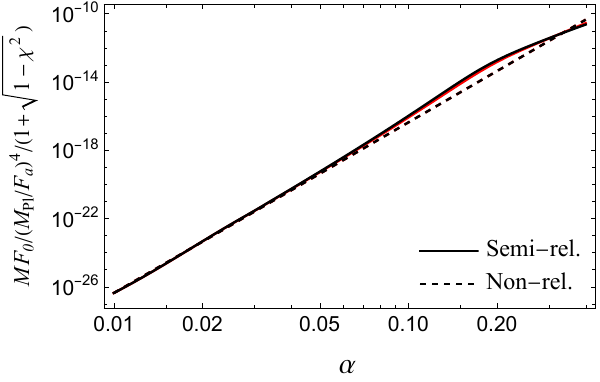}
\caption{The coefficient of the flux at the horizon induced by the self-interaction, $F_0$. Solid (Dashed) line shows the result of a semi-relativistic (non-relativistic) approximation. BH spin is $\chi=0.9$ (black) and $\chi=\chi_{\rm crit}$ (red).}
\label{fig:F0}
\end{figure}

However, we note that the non-relativistic approximation works well for axions at the mode satisfying $\alpha/l\ll1$. 
Thus, even for the region of $\alpha$ where $l=m=1$ mode is non-relativistic, $l=m=0$ mode can be relativistic. 
In this paper, to remedy the drawback of the non-relativistic approximation, we use the relativistic results calculated numerically for the eigenfrequency and the radial wavefunction of $l=m=0$ mode, referring to this hybrid method as {\it semi-relativistic approximation}. In Fig.~\ref{fig:F0}, we show $F_0$ that we use in this paper instead of Eq.~\eqref{eq:f0nr} or Eq.~\eqref{eq:F0nr}. For a detailed method of calculation, see Ref.~\cite{Omiya:2022gwu}, for example.

Indeed, $F_0$ calculated in the semi-relativistic approximation can become more than an magnitude larger compared to one calculated in the non-relativistic approximation, around $\alpha\simeq0.15$.
This implies that that the energy transfer from the primary cloud to the secondary cloud is underestimated in the non-relativistic calculation.
Therefore, this improvement will be significant for the calculation of the cloud mass in Sec.~\ref{sec:results}.

\subsection{Tidal interaction}
The second term on the right-hand side of Eq.~\eqref{eq:eom}, the tidal interaction,
induces the transition to another mode and unbound modes.
For a binary companion with mass $M_{\ast}$ at the position $(R_{\ast},\Theta_{\ast},\Phi_{\ast})$, the tidal potential is given by
\footnote{As mentioned in Ref.~\cite{Tomaselli:2023ysb}, there is a $l_{\ast}=1$ component outside the companion.
However, in the evolution of the binary system, the dissipation of axions to infinity begins to be contributed more quickly by larger $l_\ast$ components. If the mass ratio is not too small, the cloud will dissipate through the mode excited by $l_\ast\geq2$ components. Thus, we omit the $l_\ast=1$ component here.}
\begin{align}\label{eq:tidal_potenital}
    V_{\ast}=-q\alpha\sum_{l_{\ast}\leq2}\sum_{|m_{\ast}|\leq l_{\ast}}\frac{4\pi}{2l_{\ast}+1}\frac{r_{<}^{l_{\ast}}}{r_{>}^{l_{\ast}+1}}Y^{\ast}_{l_{\ast}m_{\ast}}(\hat{\bm{n}}_{\ast})Y_{l_{\ast}m_{\ast}}(\hat{\bm{n}})~,
\end{align}
where the mass ratio is defined as $q=M_{\ast}/M$, $r_{>}$ ($r_{<}$) is the larger (smaller) of $r$ and $R_{\ast}$, and $\hat{\bm{n}}=(\theta,\varphi)$, $\hat{\bm{n}}_{\ast}=(\Theta_{\ast},\Phi_{\ast})$.
For simplicity, we consider the orbit is on the equatorial plane ($\Theta_{\ast}=\pi/2$), and quasi-circular.
Orbital angular velocity is defined as
\begin{align}
    \Omega(t)=\pm\dot{\Phi}_{\ast}~,
\end{align}
where the upper (lower) sign represents the case of co-rotating (counter-rotating) orbits.
It is related to the separation through Kepler's law as
\begin{align}
    \Omega^2=\frac{M}{R_\ast^3}(1+q)~.
\end{align}
For clean binary systems, $\Omega^2$ gradually increases due to the radiation reaction of GW emission as~\cite{PhysRev.136.B1224, Blanchet:2013haa}
\begin{align}
    \frac{d\Omega}{dt}&=\gamma\left(\frac{\Omega}{\Omega_0}\right)^{11/3}~, \label{eq:orbital} \\
    \gamma&=\frac{96}{5}\frac{q}{(1+q)^{1/3}}(M\Omega_0)^{5/3}\Omega_0^2~,
\end{align}
where $\Omega_0$ is a reference value.
The timescale of the binary evolution at $\Omega=\Omega_0$ is given by
\begin{align}
    \tau_{\rm bin}=\frac{\Omega_0}{\gamma}~. \label{eq:tau_bin}
\end{align}

From now on, we discuss the dissipation of the axion cloud composed of superradiant $\ket{211}$ or $\ket{322}$ states induced by the tidal interaction.
Note that the ratio of the tidal potential to the gravitational potential to the central BH is $V_\ast/(\alpha/r)\sim q(r_c/R_\ast)^3$, where the cloud radius is estimated as $r_c\sim n^2/\alpha^2$.
Thus, when $R_\ast\gg r_c$, the perturbative approach should work even for $q\sim1$, except near the resonacne (see App.~\ref{app:two}).

\subsubsection{Axion emission and absorption}
When the orbital angular velocity becomes larger than the value that excites axions to unbound modes, axions are emitted away to infinity. 
In addition, tidal interaction can induce the transition to the mode with a large decay rate that is absorbed by the BH directly.
In calculating the flux, we approximate the orbital angular velocity as a constant during the evolution of the cloud.
Therefore, the flux ${\cal F}^{\rm td}$ induced by tidal interaction for each $(l_{\ast},m_{\ast})$ component can be
calculated with the source
\begin{align}
    {\cal S}=V_{l_{\ast},m_{\ast}}e^{\mp i m_{\ast}\Omega t}\psi_i~,
\end{align}
where $V_{l_{\ast},m_{\ast}}$ is the $(l_{\ast},m_{\ast})$ component of Eq.~\eqref{eq:tidal_potenital} excluding the time-dependence.

For instance, if $|m_{\ast}|\Omega>\mu-\omega_{i,R}$, 
the flux to infinity induced by the $(l_{\ast},m_{\ast})$ component has a finite non-vanishing value.
The flux at infinity for axions at $i$ state is obtained by Eq.~\eqref{eq:FI2}.
Explicit expression is
\begin{align}
    {\cal F}_{i,{\cal I}}^{\rm td}=\sum_{l_{\ast},m_{\ast}}\sum_{l,m}\frac{4\mu k}{\left|W_l(\omega)\right|^2}\left.r^2|R_{kl}^{\rm up}|^2\right|_{r\to\infty}\eta_{\cal I}^2(k)\Theta(k^2) N_i~,
\end{align}
with
\begin{align}
    \eta_{\cal I}(k)&=\left|\int d^3x R_{kl}^{\rm in}Y_{lm}^{\ast}V_{l_{\ast}m_{\ast}}R_{n_il_i}Y_{l_im_i}\right|~, \\
    \omega&=\omega_i\pm m_{\ast}\Omega~, \\
    k&=\sqrt{2\mu(\omega_i\pm m_{\ast}\Omega-\mu)}~,
\end{align}
where $\Theta$ is the Heaviside step function. The in-mode and the up-mode wavefunctions are given by Eq.~\eqref{eq:in_mode} and Eq.~\eqref{eq:up_mode}, respectively.
Here, $i$ is the label for the states $\ket{211}$ ($i=1$) and $\ket{322}$ ($i=2$).

Similarly, the flux at the horizon carried by the mode induced by the tidal interaction can be calculated~\footnote{In Ref.~\cite{Takahashi:2023flk}, we used the induced decay rate of the $\ket{211}$ state due to the coupling with $\ket{21\,\mbox{-1}}$ calculated by using the two-mode approximation as discussed in App.~\ref{app:two}. However, we confirmed that the two-mode approximation agrees well with that calculated by using the perturbation theory described in the main text when the system is near the resonance.}.
It is given by Eq.~\eqref{eq:FH2}. Explicitly,
\begin{align}
    {\cal F}_{i,{\cal H}}^{\rm td}=\sum_{l_\ast m_\ast}\sum_{lm}8Mr_+\mu \frac{\mu-m\Omega_H}{\left|W_l(\omega)\right|^2}\left|R_{kl}^{\rm in}(0)\right|^2\eta_{\cal H}^2(k)N_i~,
\end{align}
with
\begin{align}
    \eta_{\cal H}(k)&=\left|\int d^3x R_{kl}^{\rm up}Y_{lm}^{\ast}V_{l_{\ast}m_{\ast}}R_{n_il_i}Y_{l_im_i}\right|~.
\end{align}
This process is especially relevant when the $l=m=0$ mode is induced.
For instance, $\ket{322}$ state can couple to the $l=m=0$ mode through the $(l_{\ast},m_{\ast})=(2,-2)$ component
of the tidal perturbation for the co-rotating case.
As in the case of self-interaction, the flux for this process is calculated using the semi-relativistic approximation.

\subsubsection{Indirect emission}\label{subsubsec:indirect}

During the inspiral phase, the orbital frequency also experiences the resonance frequencies between the eigenmodes of the axion.
As discussed above, when the companion approaches sufficiently close, the eigenmode becomes unstable due to the emission to infinity.
Even if the orbital frequency is not large enough to directly excite the superradiant mode of interest to unbound modes, higher energy modes can be excited.
Then, there is a process in which axions transferred to the higher energy modes are subsequently excited to unbound modes.
For instance, this process proceeds as  $\ket{211}\to$ the bound mode with $l=m=3\to$ the unbound mode with $l=m=5$.
We can calculate the flux of the indirect emission through this process as the second order perturbation, including the resonant transitions.

Similar to the first order solution in Eq.~\eqref{eq:pertsol}, the second order solution is constructed using the Green's function in Eq.~\eqref{eq:Green} as
\begin{align}
    \psi^{(2)}=\int d^4\! x'\ G(x,x')V_{l_{\ast},m_{\ast}}e^{\mp i m_{\ast}\Omega t}\psi^{(1)}~.
\end{align}
In this process, we consider the case where the angular frequency of the first order solution is smaller than $\mu$ such that it is a bound mode.
The flux at infinity is given by
\begin{align}
    {\cal F}_{i,{\rm Ind}}^{\rm td}=\sum_{l_{\ast}^{(2)}m_{\ast}^{(2)}}\sum_{l^{(2)}m^{(2)}}&\frac{4\mu k_2}{\left|W_{l^{(2)}}(\omega_2)\right|^2}\left.r^2|R_{k_2l^{(2)}}^{\rm up}|^2\right|_{r\to\infty} \notag \\
    &\times\eta_{{\cal I},(1)}^2(k_2)\Theta(k_2^2) N_i~, \label{eq:Find}
\end{align}
where
\begin{align}
    &\eta_{{\cal I},(1)}(k_2) \notag \\
    &=\sum_{l_\ast m_\ast}\sum_{lm}
    \left|\int d^3x R_{k_2l^{(2)}}^{\rm in}Y_{l^{(2)}m^{(2)}}^{\ast}V_{l_{\ast}^{(2)}m_{\ast}^{(2)}}R_{k_1l}^{(1)}Y_{lm}\right|~,
\end{align}
\begin{align}
    &R_{k_11}^{(1)}(r)=\frac{2\mu}{W_l(\omega_1)} \notag \\
    &\times\left[R_{k_1l}^{\rm up}(r)\int_0^r dr'\int d\cos\theta d\varphi R_{k_1l}^{\rm in}Y_{lm}^\ast V_{l_\ast m_\ast} R_{n_il_i}Y_{l_im_i}\right. \notag \\
    &+\left.R_{k_1l}^{\rm in}(r)\int_r^\infty dr'\int d\cos\theta d\varphi R_{k_1l}^{\rm in}Y_{lm}^\ast V_{l_\ast m_\ast} R_{n_il_i}Y_{l_im_i}\right]~,
\end{align}
with
\begin{align}
    \omega_1=\omega_i\pm m_\ast\Omega~,\quad k_1=\sqrt{2\mu(\omega_1-\mu)}~, \\
    \omega_2=\omega_1\pm m_\ast^{(2)}\Omega~,\quad k_2=\sqrt{2\mu(\omega_2-\mu)}~.
\end{align}

Note that, regarding the calculation of the indirect emission, it is necessary to pay attention to the validity range of each approximation.
While the perturbation theory breaks down when the frequency of the intermediate state is near the resonance, the two-mode approximation should be valid there.
On the other hand, when the intermediate state is off-resonance, the two-mode approximation can be very wrong.  
We compare these two approximations in App.~\ref{app:two}.
However, we will see later that including the  indirect emission does not significantly change the evolution.

Combining the above argument, the flux of the axion dissipated by the tidal interaction is given by
\begin{align}
    {\cal F}^{\rm td}_i={\cal F}^{\rm td}_{i,{\cal I}}+{\cal F}^{\rm td}_{i,{\cal H}}+{\cal F}^{\rm td}_{i,{\rm Ind}}~. \label{eq:Ftidal}
\end{align}

\section{Evolution equations}\label{sec:evolution_eq}
In this section, we give the equations that describe the system's evolution.
First, the evolution of the cloud is calculated using the net flux discussed in the previous section.
Equations for the particle numbers at 
the states $\ket{211}$ (labeled by 1) and $\ket{322}$ (labeled by 2) are
\begin{align}
    \frac{dN_1}{dt}=2\omega_{1,I}N_1-{\cal F}^{\rm SI}_1-{\cal F}^{\rm td}_1~, \label{eq:N1} \\
    \frac{dN_2}{dt}=2\omega_{2,I}N_2-{\cal F}^{\rm SI}_2-{\cal F}^{\rm td}_2~, \label{eq:N2}
\end{align}
where ${\cal F}^{\rm SI}_i$ are given by Eqs.~\eqref{eq:Fself1} and \eqref{eq:Fself2}, and ${\cal F}^{\rm td}_i$ are given by Eq.~\eqref{eq:Ftidal}.

As the cloud grows, angular momentum is extracted from the central BH. 
The angular momentum flux for each mode is related to the energy flux as $\dot{J}_{{\rm c},i}\simeq(m/\mu)\dot{M}_{{\rm c},i}$.
Thus, the local angular momentum conservation at the horizon reads
\begin{align}
    \frac{dJ}{dt}=-\frac{2\omega_{1,I}}{\mu}M_{{\rm c},1}-\frac{4\omega_{2,I}}{\mu}M_{{\rm c},2}~,
\end{align}
where the cloud mass is given by $M_{{\rm c},i}=\mu N_i$.
Here, we ignored the flux through the other induced modes
\footnote{As we studied in Ref.~\cite{Takahashi:2023flk}, when we focus on the hyperfine resonance between $\ket{211}$ and $\ket{21\,\mbox{-1}}$, the flux of the transition destination mode is also important if the cloud mass is large and the BH spin is saturated at $\chi_{\rm crit}$. Now, we consider the case where the BH spin is not completely saturated. In addition, since other induced modes absorbed by the BH have $m=0$, they do not contribute to the change of the angular momentum.}.
We also ignore the change of the BH mass since the cloud mass is small when the self-interaction works efficiently.
For cases where the local energy conservation at the horizon is considered, see the Ref.~\cite{Takahashi:2023flk}.

In addition, the orbital motion of the binary is affected by the presence of the cloud.
The tidal interaction transfers the angular momentum between the clouds and the orbital motion.
Here, we decompose the number fluxes induced by the tidal interaction ${\cal F}_i^{\rm td}$ as
\begin{align}
    {\cal F}_i^{\rm td}=\sum_m {\cal F}_{i,m}^{\rm td}~,
\end{align}
where ${\cal F}_{i,m}^{\rm td}$ represent the components due to the transition to a dissipative mode with the magnetic quantum number $m$.
From the total angular momentum conservation, a term representing the angular momentum transfer is added to Eq.~\eqref{eq:orbital} as
\begin{align}
    \frac{d\Omega}{dt}=\gamma\left(\frac{\Omega}{\Omega_0}\right)^{11/3}&\pm R\left(\frac{\Omega}{\Omega_0}\right)^{4/3}\frac{\Omega_0}{M^2} \notag \\
    &\times\sum_m \left[(m-1){\cal F}^{\rm td}_{1,m}+(m-2){\cal F}^{\rm td}_{2,m}\right]~,
\end{align}
where $R=3(1+q)^{1/3}q^{-1}(M\Omega_0)^{1/3}$.
Now, we can solve the evolution of the system of our interest.
In practical calculations, we consider only the dominant processes for each phenomenon of interest.

\section{Phenomenology}\label{sec:results}
Finally, we show the phenomenological consequences of the self-interacting axion clouds in binary systems during the inspiral phase.

\begin{figure*}[t]
\includegraphics[keepaspectratio, scale=0.7]{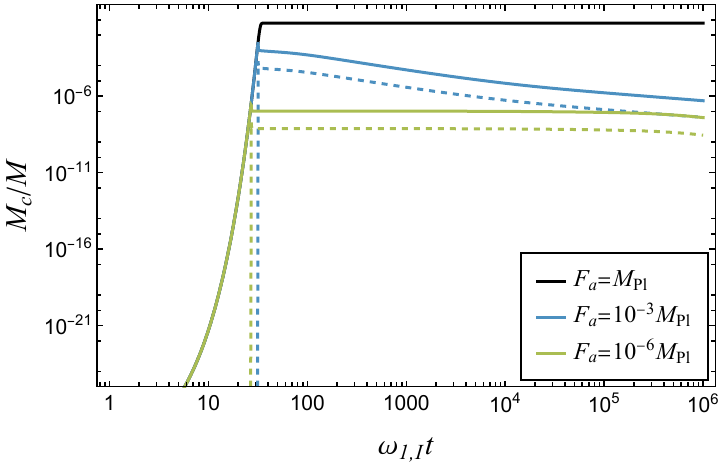}
\includegraphics[keepaspectratio, scale=0.7]{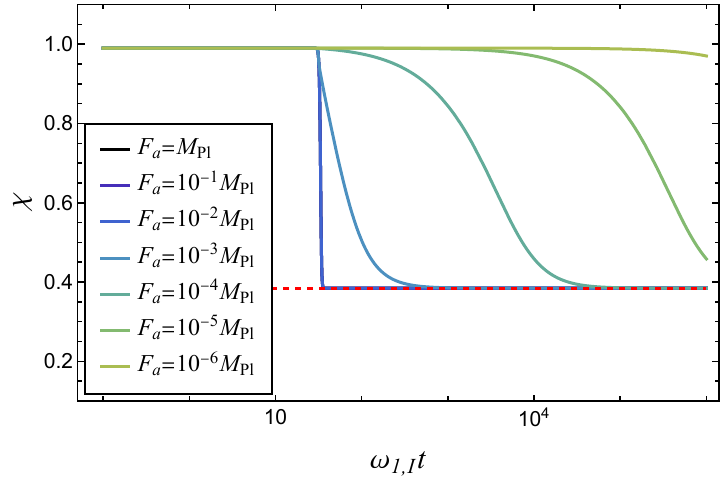}
\caption{\label{fig:evolution_self} Evolution of the cloud mass (left) and the BH spin (right) when we consider only the superradiance and self-interaction. Time is normalized by the superradiant growth time scale of the primary cloud $1/\omega_{1,I}$ at the initial BH spin. In the left panel, solid (dashed) curves show the mass of the cloud composed of axions at  $\ket{211}$ ($\ket{322}$) for various $F_a$. In the right panel, the red dashed line shows the critical spin value $\chi_{\rm crit}$. The initial spin is given by $\chi=0.99$ and we set $\alpha=0.1$. The curves for $F_a=10^{-2}M_{\rm Pl}, 10^{-1}M_{\rm Pl}$ and $M_{\rm Pl}$ are almost on top of each other.}
\end{figure*}

\subsection{Lifetime of the BH spin}

\begin{figure*}[t]
\includegraphics[keepaspectratio, scale=0.85]{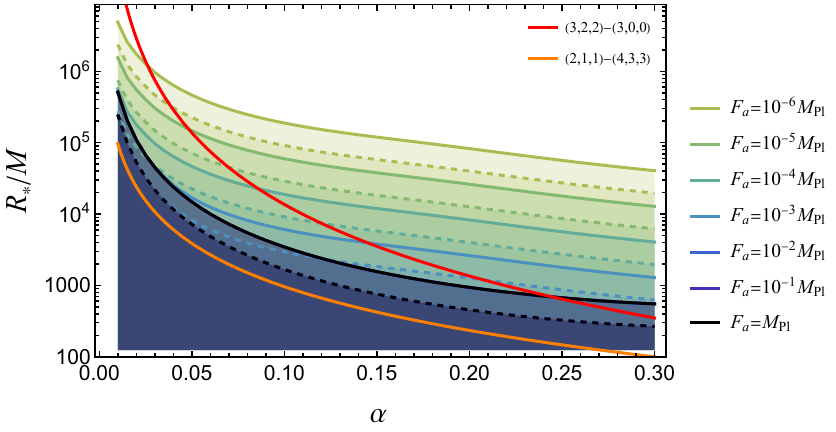}
\caption{\label{fig:lifetime}  Boundaries of the regions where the timescale of the binary evolution is shorter than the lifetime of the BH spin for various $F_a$. In the colored region, $\tau_{\rm bin}<\tau_{\rm lifetime}$ holds, where $\tau_{\rm lifetime}$ is the time for the BH spin to reach $\chi=\chi_{\rm crit}$ from $\chi=0.99$. 
The solid curves are for $q=1$, and the dashed ones are for $q=0.1$. 
The red and orange curves show the separation at the resonance between $\ket{322}$ and $\ket{300}$, and that between $\ket{211}$ and $\ket{433}$, respectively. }
\end{figure*}

If the binary system is formed at a sufficiently large separation where the effects of the tidal resonance and dissipation can be neglected, the evolution of the BH-cloud is determined only by the superradiance and the self-interaction at the early stage.
In this case, evolution equations for particle numbers are given by
\begin{align}
    \dot{N}_1&=2\omega_{1,I}N_1-2f_0N_1^2N_2+f_3N_1N_2^2~,\\
    \dot{N}_2&=2\omega_{2,I}N_2+f_0N_1^2N_2-2f_3N_1N_2^2~. \label{eq:N2self}
\end{align}
We show an example of the evolution of the cloud mass and the BH spin in Fig.~\ref{fig:evolution_self} for $\alpha=0.1$ and various $F_a$ with the initial BH spin $\chi=0.99$.
When the decay constant is small (or when the self-interaction is strong), the primary cloud remains small. This is because the energy transfer to the secondary cloud, associated with the absorption by the horizon, hinders the growth of the primary cloud. 
After that, the reverse energy transfer from the secondary to the primary becomes important, and a quasi-stationary state with a stable flow of energy and angular momentum from the BH to infinity is achieved.
As the BH angular momentum continues to be extracted, the cloud masses gradually decrease.
Compared with the timescale of BH spin-down, the change of the cloud mass is slow.
Thus, we can obtain each cloud mass by solving $\dot{N}_1\simeq 0$ and $\dot{N}_2\simeq 0$ for a given BH spin approximately. Neglecting $2\omega_{2,I}$ in Eq.~\eqref{eq:N2self}, we get the saturated cloud masses (normalized by the BH mass) as
\begin{align}\label{eq:saturated_mass}
    \tilde{M}_{{\rm c},1}^{\rm s}\simeq\sqrt{\frac{8F_3}{3F_0^2}\omega_{1,I}}~, \quad \tilde{M}_{{\rm c},2}^{\rm s}\simeq\sqrt{\frac{2}{3F_3}\omega_{1,I}}~.
\end{align}

When the decay constant is small, the BH spin-down becomes slow.
Therefore, the lifetime of the BH spin becomes much longer than that of the non-interacting case.
Depending on the formation process of binary systems, the tidal interaction can start working before the BH spin decays down to the threshold value of the superradiance.
Such cases are interesting because the tidal interaction comes into play when the clouds are in the quasi-stationary state. 
To realize this situation, the timescale of binary evolution in Eq.~\eqref{eq:tau_bin} at the formation must be shorter than the lifetime of the BH spin $\tau_{\rm lifetime}$, {\it i.e.}, 
\begin{align}
    \tau_{\rm bin}<\tau_{\rm lifetime}~.
\end{align}
Here, we evaluate $\tau_{\rm lifetime}$ as the time for the BH spin to reach $\chi=\chi_{\rm crit}+10^{-2}$ from $\chi=0.99$, obtained by solving the evolution equations.
In Fig.~\ref{fig:lifetime}, we show the binary separation $R_\ast$ that satisfy $\tau_{\rm bin}=\tau_{\rm lifetime}$ as a function of $\alpha$ for various $F_a$.
A binary system formed at a separation smaller than this boundary is affected by the tidal interaction while the BH spin remains.
In the same figure, the separations corresponding to typical resonance frequencies are also shown.
Since the timescale of the binary evolution can change due to orbital sink or float caused by the tidal resonance and dissipation~\cite{Baumann:2019ztm,Takahashi:2023flk}, this is a naive condition.

Note that, if there is too much time after the termination of the superradiance of the primary cloud due to the BH spin-down, the secondary cloud or the cloud composed of modes with $l \ge 3$ begins to grow~\cite{Omiya:2022gwu}.
In this paper we focus on the case in which the $\ket{211}$ cloud dominates, supported by the superradiance.

\subsection{Tidal disruption and dephasing of GWs}

\begin{figure*}[t]
\includegraphics[keepaspectratio, scale=0.75]{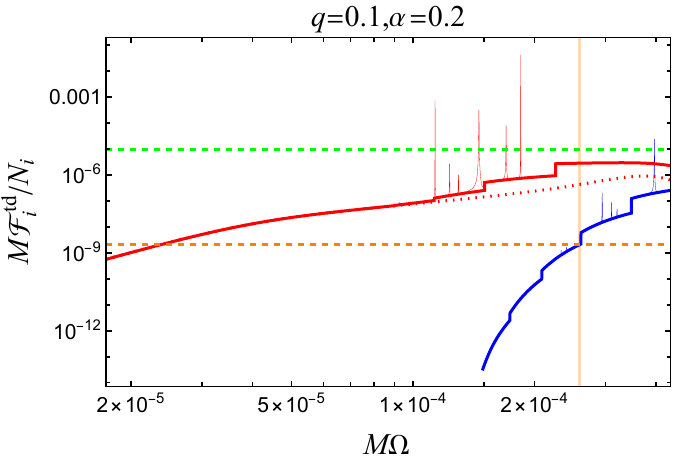}
\includegraphics[keepaspectratio, scale=0.75]{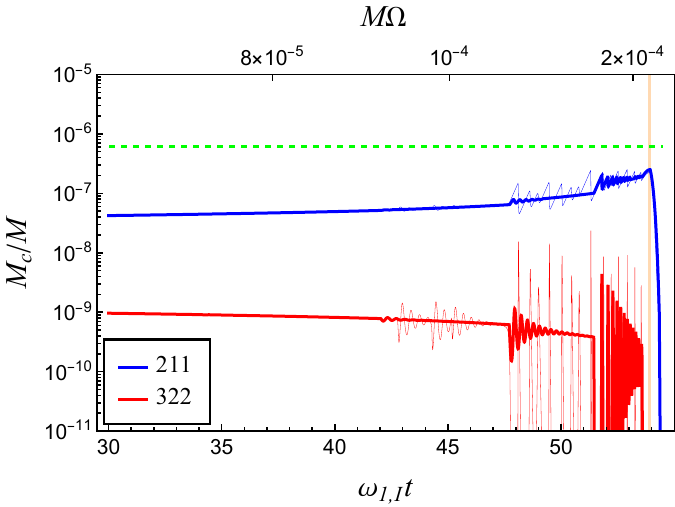}
\caption{The fluxes induced by the tidal interaction (left) and the evolution of the cloud mass (right) for $q=0.1,\alpha=0.2$, and $F_a=10^{-5}M_{\rm Pl}$. In the left panel, the blue (red) solid curve shows the fluxes for the $\ket{211}$ ($\ket{322}$) cloud. The red dotted curve shows ${\cal F}_{2,{\cal H}}^{\rm td}$. The horizontal orange dashed line shows the superradiant growth rate $2\omega_{1,I}$ for $\chi=0.9$. In the right panel, the blue (red) solid curve shows the time evolution of the $\ket{211}$ ($\ket{322}$) cloud mass normalized by the central BH mass. They are obtained by solving the equations discussed in Sec.~\ref{sec:evolution_eq} with initial conditions $M_{{\rm c},i}/M=\tilde{M}_{{\rm c},i}^{\rm s}$, $\chi=0.9$, and sufficiently small orbital frequency. In both panels, the curves drawn by thick and thin lines represent the cases with and without the indirect emission. The orange vertical lines show the orbital frequency and time when ${\cal F}_1^{\rm td}/N_1$ equals to $2\omega_{1,I}$. Green dashed lines will be explained in Fig~.{\ref{fig:evolution_BN}} and Sec.~\ref{subsubsec:BN}.}
\label{fig:evolution_TD}
\end{figure*}

From now on, we consider the tidal interaction that begins to act effectively as the orbital frequency increases during the inspiral.
We assume that the orbit is co-rotating to the spin of the central BH and the clouds are in a quasi-stationary state.
First, we show an example of the time evolution for $q=0.1$, $\alpha=0.2$ and $F_a=10^{-5}M_{\rm Pl}$ in Fig.~\ref{fig:evolution_TD}.
In the left panel, the particle number fluxes induced by the tidal interaction ${\cal F}_i^{\rm td}$ as a function of the orbital velocity for $\ket{211}$ and $\ket{322}$ clouds are shown. 
As one can see, the flux for $\ket{322}$ is larger than that for $\ket{211}$.
At the low $\Omega$ region, the flux for $\ket{322}$, ${\cal F}_2^{\rm td}$, is dominated by the flux to the horizon ${\cal F}_{2,{\cal H}}^{\rm td}$, which is induced by the coupling with $l=m=0$ mode
\footnote{${\cal F}_{1,{\cal H}}^{\rm td}$ also has a finite value due to the coupling with $l=1,m=-1$ mode. However, it is subdominant except for the resonance in the current situation. Even at the resonance, its effect is negligible when the mass ratio is large enough~\cite{Takahashi:2021yhy}} (see dotted line in Fig.~\ref{fig:evolution_TD}).
As $\Omega$ increases, the fluxes escaping to infinity, ${\cal F}_{2,{\cal I}}^{\rm td}$, become dominant. 
For simplicity, we consider only the fluxes of modes with $l=m$, since the contribution of the other modes are subdominant.
Also, regarding the indirect emission, we consider only the fluxes of modes excited by the $l_\ast^{(2)}=2$ component of the tidal potential as the dominant contribution and include up to the first three resonances.

As discussed in the App.~\ref{app:two}, there are subtleties in calculating the fluxes of the indirect emission near the resonance.
Therefore, 
we show the results with and without the indirect emission 
using thin and thick lines in the figure, respectively.
For the current parameters, the flux related to the transition from $\ket{322}$ 
 to an unbound mode with $l=m=6$ first exceeds ${\cal F}_{2,{\cal H}}^{\rm td}$ at $M\Omega\simeq 1.2\times10^{-4}$.
 Several spikes drawn with thin lines represent the contribution of the indirect emission near the resonances
 \footnote{In the non-relativistic approximation, the indirect flux formally diverges since the Wronskian in Eq.~\eqref{eq:Find} vanishes at the resonance. Thus, the peak amplitude shown in the left panel of Fig.~\ref{fig:evolution_TD} is artificial.}.
 For instance, the transition from $\ket{322}$ to a bound mode with $l=m=5$ and the successive one to an unbound mode with $l=m=7$ contribute at a similar $\Omega$.
Regarding $\ket{211}$ cloud, the flux ${\cal F}_1^{\rm td}/N_1$ first exceeds the superradiant growth rate $2\omega_{1,I}$ for $\chi=0.9$ at $M\Omega\simeq 2.6\times10^{-4}$, which is shown as the orange vertical line in the left panel of Fig.~\ref{fig:evolution_TD}.
The dominant contribution at this point is the flux related to the transition from $\ket{211}$ to an unbound mode with $l=m=6$.
As $\Omega$ increases, the direct transition to an unbound mode with smaller $l$ becomes dominant.

\begin{figure}[t]
\includegraphics[scale=0.8]{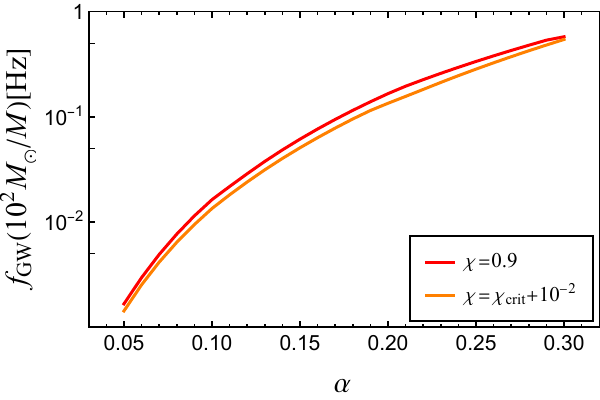}
\caption{GW frequency where the cloud disappears for $q=0.1$. The red line and orange line represent the GW frequency where the escaping flux induced by the tidal interaction for $\ket{211}$ cloud, ${\cal F}_1^{\rm td}/N_1$, equals to the superradiant growth rate $2\omega_{1,I}$ for $\chi=0.9$ and $\chi=\chi_{\rm crit}+10^{-2}$, respectively.}
\label{fig:fgw}
\end{figure}

In the right panel, we show the time evolution of the cloud masses.
As the initial conditions, we take $M_{{\rm c},i}/M=\tilde{M}_{{\rm c},i}^{\rm s}$ given in Eq.~\eqref{eq:saturated_mass}, $\chi=0.9$ and $\Omega$ is set to $0.1$ times the orbital frequency at which the $l_\ast=2$ component of the tidal potential begins to excite the $\ket{211}$ mode to an unbound mode.
Since the $\ket{322}$ cloud is supported by the energy transfer from the $\ket{211}$ cloud, which is supported by the superradiance, once the ${\cal F}_2^{\rm td}$ exceeds the superradiant growth rate $2\omega_{i,I}$, the mass of the $\ket{322}$ cloud begins to decrease.
Then, the energy transfer from the $\ket{211}$ cloud to the $\ket{322}$ cloud will be suppressed, and the mass of the $\ket{211}$ cloud will slightly increase to balance.
When the orbital frequency reaches a frequency where ${\cal F}_1^{\rm td}$ exceeds $2\omega_{1,I}$, the mass of the $\ket{211}$ cloud also begins to decrease.
Except at around the resonance frequencies, the flux induced by the tidal interaction increases as $\Omega$ increases.
Thus, both clouds disappear as a consequence of the tidal disruption.
The orbital frequency at the corresponding time is shown on the upper axis, and the orange vertical line is the same as the one in the left panel.

To find the signature of the cloud from GW observations, it is important to know when clouds disappear during the inspiral.
In Fig.~\ref{fig:fgw}, we show the GW frequency where the flux for $\ket{211}$ cloud ${\cal F}_1^{\rm td}$ crosses the superradiant growth rate $2\omega_{1,I}$ for $\chi=0.9$ and $\chi=\chi_{\rm crit}+10^{-2}$ as a function of $\alpha$ in the case of $q=0.1$.
Smaller values of $\alpha$ correspond to a larger spatial extension of the cloud, making it more susceptible to tidal interaction with lower orbital frequency.
These frequencies can be regarded as a threshold for the GW frequency below which the cloud can remain.

\begin{figure}[t]
\includegraphics[scale=0.65]{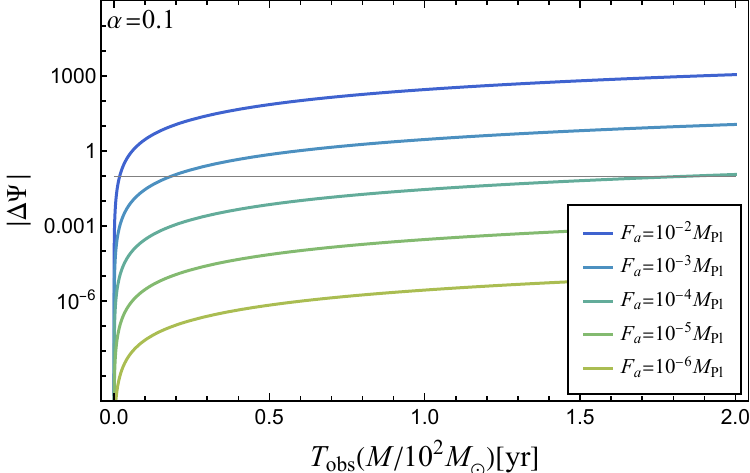}
\includegraphics[scale=0.65]{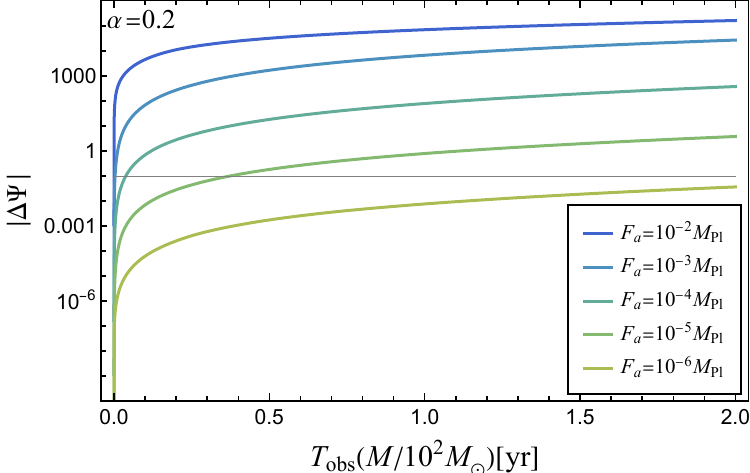}
\caption{Dephasing of the GWs due to the presence of the clouds as a function of the observation time for $q=0.1,\alpha=0.1$ (top) and $\alpha=0.2$ (bottom). Each color corresponds to a different $F_a$. The start of the observation is set at the frequency that reaches the resonance frequency between $\ket{322}$ and $\ket{300}$ in one year when there are no clouds.
Also, the initial BH spin is set to $\chi=\chi_{\rm crit}+10^{-2}$. In the region of $F_a$ where the cloud masses are small and the BH spin-down can be ignored, it is almost proportional to $F_a^2$.}
\label{fig:dephasing_res}
\end{figure}

\begin{figure}[t]
\includegraphics[scale=0.65]{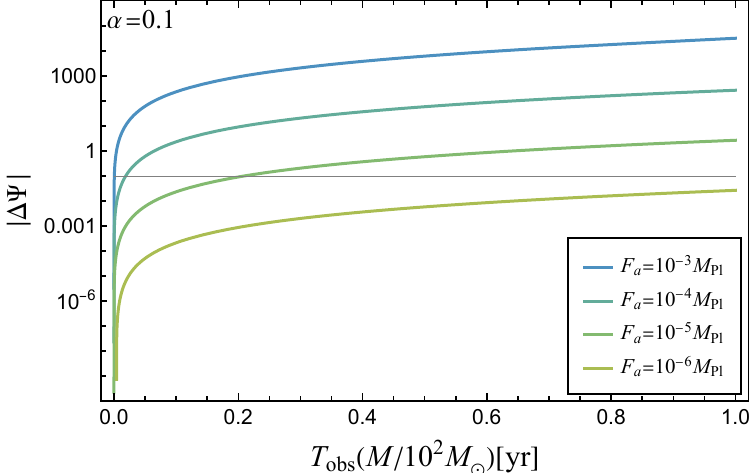}
\includegraphics[scale=0.65]{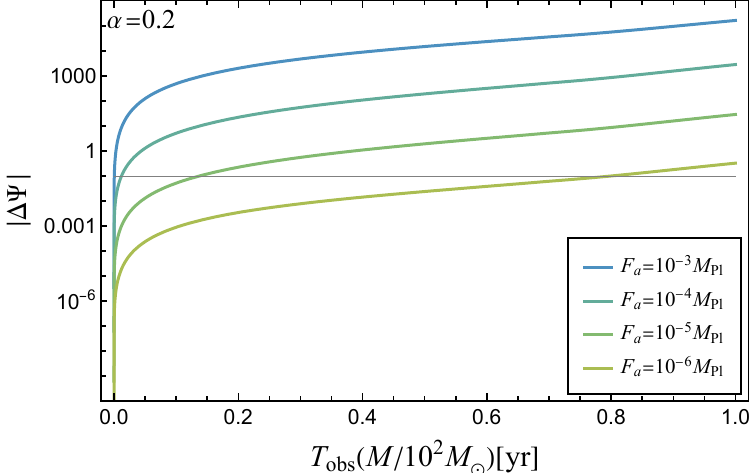}
\caption{The same figures as Fig.~\ref{fig:dephasing_res}, but the start of the observation is set as the frequency that reaches the frequency at which the clouds begin to be disrupted (shown in Fig.~\ref{fig:fgw}) in one year.}
\label{fig:dephasing_TD}
\end{figure}

While the clouds are affected by the tidal interaction, the orbital evolution also changes.
To quantify the influence of the clouds on the possible GW signal, we evaluate the dephasing of quadrupole frequencies defined as
\begin{align}
    \Delta\Psi=2\int_{0}^{T_{\rm obs}}dt\Delta\Omega(t)~.
\end{align}
Here, $\Delta\Omega(t)=\Omega(t)-\Omega_{\rm clean}(t)$ where $\Omega(t)$ is the solution obtained in our current formulation and $\Omega_{\rm clean}(t)$ is the orbital frequency in a clean binary system.
It is defined as a function of the observation period $T_{\rm obs}$.
Here, we show the results for two initial orbital frequencies. 
The first one takes the initial orbital frequency such that the resonance frequency between $\ket{322}$ and $\ket{300}$ is reached in one year, when the cloud is absent.
This resonance frequency is 
\begin{align}
    f_{\rm res}=5.7\times 10^{-4}{\rm Hz}\left(\frac{\alpha}{0.1}\right)^5\left(\frac{10^2M_{\odot}}{M}\right)~.
\end{align}
The second one sets the initial orbital frequency such that the primary cloud begins to be disrupted in one year. The GW frequency at which the disruption starts is as shown in Fig.~\ref{fig:fgw}.
In particular, for $\alpha=0.1 (0.2)$ and $\chi=\chi_{\rm crit}+10^{-2}$, the frequency is $f\simeq 0.013 (0.13) (10^2M_{\odot}/M)$ Hz.

In both cases, we adopt $\chi=\chi_{\rm crit}+10^{-2}$ as the initial BH spin.
Also, the initial cloud masses are set to the masses that balance the net flux at the initial frequency on the right-hand sides of Eqs.~\eqref{eq:N1} and \eqref{eq:N2}.
As the upper limit for the order of the magnitude of $F_a$, we chose $F_a$ such that the initial cloud masses do not exceed the central BH mass.

The results for the first and the second cases are shown in Fig.~\ref{fig:dephasing_res} and Fig.~\ref{fig:dephasing_TD}, respectively.
Importantly, considering only the self-interaction, the mass of the cloud at the quasi-stationary state is proportional to $F_a^{^2}$.
Therefore, the dephasing $|\Delta \Psi|$ magnifies as $F_a$ increases.
In the first case, $|\Delta\Psi|$ can be larger than ${\cal O}(0.1)$ in two-years observation for $F_a\gtrsim 10^{-4}M_{\rm Pl}$ when $\alpha=0.1$, and for $F_a\gtrsim 10^{-5}M_{\rm Pl}$ when $\alpha=0.2$.
In the second case, $|\Delta\Psi|$ can be larger than ${\cal O}(0.1)$ in one-year observation for $F_a\gtrsim 10^{-5}M_{\rm Pl}$ when $\alpha=0.1$, and for $F_a\gtrsim 10^{-6}M_{\rm Pl}$ when $\alpha=0.2$.

Thus, we can expect that the presence of the clouds can be investigated by future GW detectors, such as LISA~\cite{Audley:2017drz} and DECIGO~\cite{Kawamura:2011zz}.
However, to precisely evaluate the detectability, we need to examine whether we can estimate the parameters of the axion clouds, despite the degeneracy with the binary parameters that appear in the waveform without clouds.
We leave detailed analyses for future work.

\begin{figure*}[t]
\includegraphics[keepaspectratio, scale=0.75]{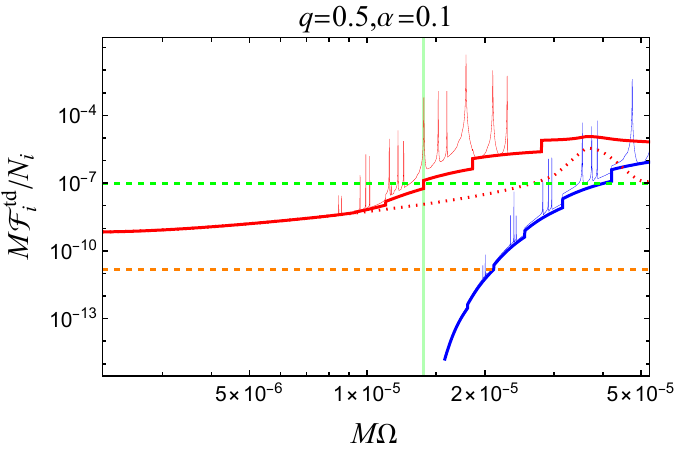}
\includegraphics[keepaspectratio, scale=0.75]{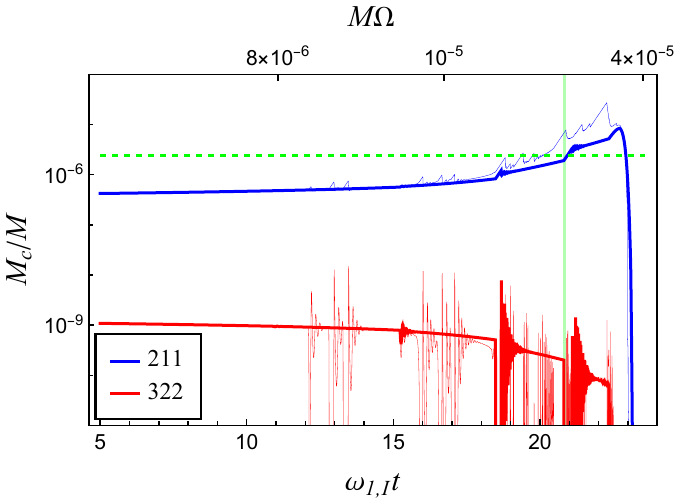}
\caption{An example of the evolution when the cloud mass exceeds the bosenova threshold: the fluxes induced by the tidal interaction (left) and the cloud mass (right) for $q=0.5,\alpha=0.1$, and $F_a=10^{-5}M_{\rm Pl}$. 
The meanings of all curves are the same as those in Fig.~\ref{fig:evolution_TD}. In the left panel, the green dashed line represents the flux induced by the self-interaction at the particle number threshold as written on the right-hand side of Eq.~\eqref{eq:BNcondition1}. In the right panel, the green dotted line represents the threshold value at which the bosenova is suggested to occur. The green vertical lines show the orbital frequency and time when ${\cal F}_2^{\rm td}/N_2$ (without the indirect emission) is equal to the threshold flux. 
After the $\ket{211}$ cloud mass exceeds the threshold, the subsequent evolution obtained by the current adiabatic treatment is meaningless, though.}
\label{fig:evolution_BN} 
\end{figure*}

\subsection{Bosenova}\label{subsubsec:BN}

One of the most interesting phenomena caused by the self-interaction to axion clouds would be the dynamical instability called ``bosenova" collapse, which is an explosion (or implosion) that occurs when the amplitude of the cloud exceeds a certain threshold.
In App.~\ref{app:BN}, we summarize the analysis of the potential stability of the axion cloud and the dynamical instability.
Under the non-relativistic approximation, the threshold value of the cloud mass for the bosenova is given by
\begin{align}
    M_{\rm BN}\simeq 246\ M \alpha^{-2}\left(\frac{F_a}{M_{\rm Pl}}\right)^2~.
\end{align}

It has been shown that when only the self-interaction effect is considered, the growth of the cloud is generally saturated before reaching the threshold~\cite{Baryakhtar:2020gao}.
However, in binary systems, the tidal interaction acts on the secondary cloud ($\ket{322}$) more efficiently than on the primary cloud ($\ket{211}$), resulting in the earlier decay of the secondary cloud.
After the secondary cloud is disrupted, the energy transfer between two clouds is suppressed and the primary cloud can grow again. 
The problem is whether the amplitude of the primary cloud can grow enough to exceed the threshold.
Below we discuss the possibility that the bosenova may occur during the binary inspiral.

For the bosenova to occur, the following conditions would be necessary.
First, to suppress the energy transfer from the primary cloud to the secondary by self-interaction, the secondary cloud must decay because of the tidal interaction. In other words, the escaping flux induced by the tidal interaction for the secondary cloud should be larger than the feed flux through the self-interaction when the primary cloud is as large as the threshold value for the bosenova,
\begin{align}\label{eq:BNcondition1}
    {\cal F}_2^{\rm td}>f_0N_{1,{\rm BN}}^2N_2~,
\end{align}
where $N_{1,{\rm BN}}=M_{\rm BN}/\mu$ is the threshold particle number for bosenova. 
In addition, the primary cloud should not be disrupted. 
Namely, in addition to the condition Eq.~\eqref{eq:BNcondition1}, we require 
\begin{align}\label{eq:BNcondition2}
    {\cal F}_1^{\rm td}<2\omega_{1,I}N_1~.
\end{align}
Besides these conditions, the conditions must be maintained for a sufficiently long time for 
the primary cloud to grow to exceed the threshold. 
Thus, the evolution equations must be solved to see whether or not the primary cloud reaches the threshold.

In Fig.~\ref{fig:evolution_BN}, we show an example of the time evolution in which the mass of the $\ket{211}$ cloud exceeds the bosenova threshold.
The parameters are $q=0.5$, $\alpha=0.1$ and $F_a=10^{-5}M_{\rm Pl}$, and the initial spin is set to $\chi=0.9$.
As is seen from the left panel, the above two conditions for fluxes are satisfied.
For the $\ket{322}$ cloud, the flux due to the direct excitation to the unbound mode with $l=m=6$ exceeds the flux related to the transfer from $\ket{211}$ to $\ket{322}$ at the threshold particle number, which is shown as the green dashed line.
This frequency is $M\Omega\simeq 1.4\times 10^{-5}$ and is shown as the green vertical lines.
Thus, the condition~\eqref{eq:BNcondition1} is satisfied.
Note that, when we include the indirect emission shown as the thin line, ${\cal F}_2^{\rm td}$ exceeds the threshold slightly earlier.
For the $\ket{211}$ cloud, the flux due to the direct excitation to the unbound mode with $l=m=7$ dominates when ${\cal F}_1^{\rm td}$ crosses $2\omega_{1,I}$ (shown by the horizontal orange dashed line). 
The region to the left of this intersection satisfies condition~\eqref{eq:BNcondition2}.
In this way, we find a region where both conditions~\eqref{eq:BNcondition1} and~\eqref{eq:BNcondition2} are satisfied.

In the right panel of Fig.~\ref{fig:evolution_BN}, we show the time evolution of the masses of the respective clouds.
In this case, the $\ket{211}$ cloud grows fast enough to reach the threshold value shown as the green dashed line. 
Thus, bosenova may occur in this case. 
The subsequent evolution becomes dynamically unstable and is beyond the scope of the current formulation.
For reference, the ratio of the maximum value obtained by solving the evolution without the indirect emission to the threshold value is $M_{{\rm c},1}^{\rm max}/M_{\rm BN}\simeq 3.4$.
Including the indirect emission, the cloud mass can reach a larger value shown as the thin line.
The maximum mass becomes $M_{{\rm c},1}^{\rm max}/M_{\rm BN}\simeq 11.2$.
While including the indirect emission helps the bosenova occur, the cloud mass can reach the threshold even within a conservative analysis that does not include the indirect emission.

\begin{figure}
\includegraphics[scale=0.7]{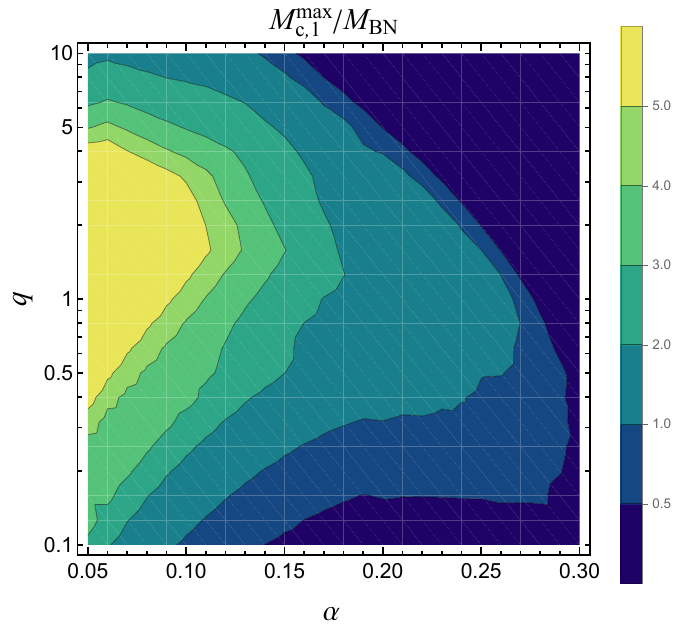}
\caption{Maximum cloud mass that can be reached under the adiabatic treatment, assuming the absence of bosenova. To be conservative, the result without the indirect emission induced by the tidal interaction is shown. The BH spin and the decay constant are set to $\chi=0.9$ and $F_a=10^{-5}M_{\rm Pl}$, respectively.}
\label{fig:region_BN}
\end{figure}

In Fig.~\ref{fig:region_BN}, we show the maximum cloud mass during the evolution in the $(\alpha,q)$ parameter space, assuming that bosenovae never happen.
Here, we neglect the indirect emission and the initial BH spin is set to $\chi=0.9$ with $F_a=10^{-5}M_{\rm Pl}$.
For this value of the decay constant, the cloud masses are small enough that the BH spin hardly changes during the evolution.
In the region where $M_{{\rm c},1}^{\rm max}/M_{\rm BN}>1$, we can expect that a bosenova occurs during the inspiral.
As $q$ increases or $\alpha$ decreases, the flux due to the tidal effect increases. 
Thus, the condition~\eqref{eq:BNcondition1} becomes easier to satisfy.
However, if $q$ is too large, the timescale of the orbital evolution becomes too fast.
Also, as $\alpha$ increases, the orbital frequency that excites the unbound modes will be large.
Thus, the cloud does not have enough time to grow, and its maximum mass becomes smaller.
Therefore, a bosenova can occur only in a limited area in this parameter space. 

\begin{figure}[t]
\includegraphics[scale=0.65]{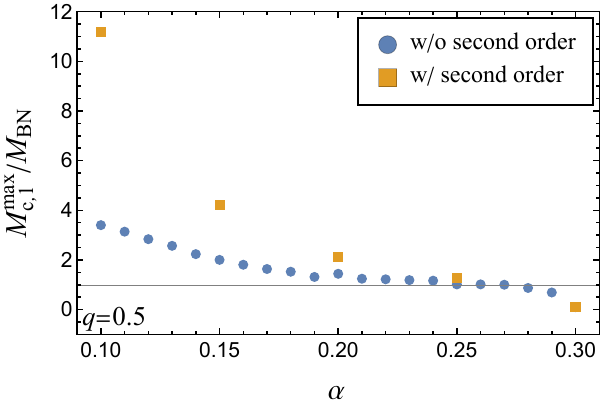}
\includegraphics[scale=0.65]{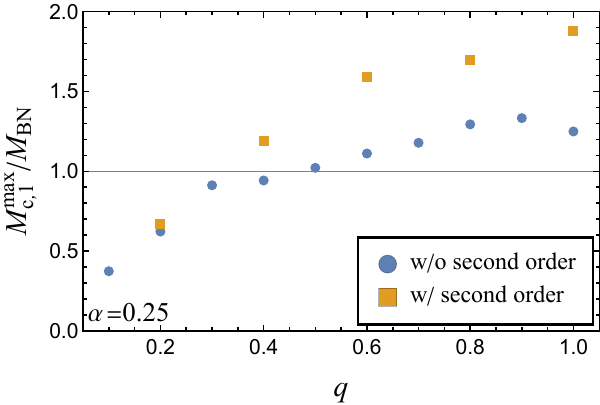}
\caption{Impact of including the indirect emission, treated by the second-order perturbation method, on the maximum cloud mass. The blue circles represent cases where the indirect emission is not included, while the yellow squares represent cases where it is included. The top panel shows results as a function of $\alpha$ with $q=0.5$. The bottom panel shows results as a function of $q$ with $\alpha=0.25$.}
\label{fig:second}
\end{figure}

As indicated in the time evolution, the indirect emission, which we include using the second-order perturbation, increases the maximum cloud mass during the evolution. 
In Fig.~\ref{fig:second}, we show the impact of the indirect emission on the maximum cloud mass.
As $\alpha$ increases, the radius of the cloud decreases, and the effect of the tidal interaction becomes smaller.
Furthermore, because the timescale of the orbital evolution becomes shorter, the impact of the indirect emission is reduced.
Therefore, we expect that the boundary where $M_{{\rm c},1}^{\rm max}/M_{\rm BN}=1$ for a smaller value of $q$ will not change significantly.
As $q$ increases, not surprisingly, the influence of the indirect emission is amplified, resulting in a larger maximum cloud mass.
Although the indirect emission significantly increases the maximum cloud mass, this is mainly due to the overestimation of this effect at resonances. Therefore, a more sophisticated treatment is required to confirm this enhancement.

Here, we fixed the decay constant to $F_a=10^{-5}M_{\rm Pl}$.
As long as $F_a$ is sufficiently small such that the BH's spin-down can be ignored, the masses of the clouds are proportional to $F_a^2$.
Larger values of $F_a$ result in more massive clouds and a stronger signal from the bosenova, such as GWs.
However, they also shorten the lifetime of the spin, reducing the likelihood of the binary separation becoming sufficiently close while the clouds remain in a quasi-stationary state.
In Fig.~\ref{fig:region_BN}, we adopt $\alpha=0.05$ as the lower limit of $\alpha$; for smaller values of $\alpha$ $M_{{\rm c},1}^{\rm max}/M_{\rm BN}$ is expected to increase.
However, the superradiant rate decreases, leading to a relatively longer timescale of the binary evolution, which makes solving the time evolution technically challenging.
Moreover, as observed in Fig.~\ref{fig:second}, the smaller the value of $\alpha$, the greater the impact of the indirect emission, worsening the perturbative treatment.
Even so, when $\alpha$ is small, the signal from the bosenova becomes weak, making this parameter region less interesting.

From a conservative standpoint, we cannot yet conclude that the bosenova can happen, since several effects are not included in our current calculation.
First, the change in the configuration of the cloud due to the self-interaction is not considered.
As the amplitude becomes larger, the attractive self-interaction makes the spatial extension of the clouds smaller~\cite{Omiya:2022gwu}.
Thus, the energy transfer from the $\ket{211}$ cloud to the $\ket{322}$ cloud is enhanced, which hinders the bosenova.
On the other hand, modification of the cloud configuration also changes the size of the flux induced by the tidal interaction.
These effects on both $\ket{211}$ and $\ket{322}$ clouds are non-trivial.
In addition, the imaginary parts of frequencies are modified by both the self-interaction and the tidal interaction.
For instance, acceleration of the instability~\cite{Omiya:2020vji} acts positively for the bosenova to occur.
Furthermore, numerical simulations will be needed 
to understand the fate of the dynamical instability.

\section{Conclusion}\label{sec:conslucion}

In this paper, we investigated the evolution of self-interacting axion clouds during the inspiral phase of binary systems. Unlike scenarios with negligible self-interaction, we need to consider two types of clouds: a primary cloud formed by the superradiance and a secondary cloud formed by the mode coupling. The tidal interaction has distinct quantitative effects on each cloud, and we also anticipate the occurrence of a dramatic phenomenon, bosenova.

Given the adiabatic nature of the growth of the axion clouds, we introduced a formulation to analyze the evolution adiabatically based on the conservation laws. We computed fluxes resulting from the self-interaction and the tidal interaction in a perturbative manner.
We examined the dominant processes between the two clouds with self-interaction based on our previous research.
The calculation of the flux to the BH for the $l=m=0$ mode was improved by treating it as a semi-relativistic approximation.
Regarding tidal interaction, we considered three dissipation processes: absorption into the BH, and the direct and indirect emissions to infinity.
The first process is selectively relevant to the secondary cloud, designated as $\ket{322}$.
The last process, which includes dissipation associated with resonant transitions, was described as a second-order perturbation.

As a result, we have identified several implications for the evolution properties and observational signatures. Only the self-interaction is presumed to be active during the early stages of cloud growth, and the axion clouds settle to a quasi-stationary state composed of the primary and secondary modes. By calculating the lifetime of the BH's spin, we have confirmed that 
the quasi-stationary state lasts sufficiently long for the orbit to shrink and 
for the tidal interaction to become effective.

Furthermore, we have illustrated the evolution of clouds when subjected to tidal disruption. As the escaping flux from the secondary cloud surpasses the superradiant growth rate of the primary cloud, the mass of the secondary starts to decrease. Conversely, the mass of the primary cloud increases, 
but the escaping flux from the primary cloud grows and eventually surpasses its growth rate due to superradiance, leading to its disappearance.
We then showed the GW frequency at which the clouds begin to disappear, confirming that the clouds can exist within the frequency band of future space-based GW detectors.
In particular, we demonstrated that even clouds with relatively small mass, corresponding to a small decay constant, can cause large modifications to the phase evolution of GWs. 

Finally, we have explored the possibility of bosenova collapse during the inspiral phase. Our findings indicate the existence of a parameter space where the mass of the primary cloud exceeds the threshold value within the adiabatic treatment.
In addition, the indirect emission using the second-order perturbation enlarges the possibility for a bosenova to occur.
However, as listed in the last paragraph of Sec.~\ref{subsubsec:BN}, drawing a conclusive statement requires further investigation of potential effects not considered in the present paper.
Future work will involve examining these effects, analyzing the dynamical instability using a more advanced method, and discussing observational signatures.

\begin{acknowledgments}
We thank Hui-Yu Zhu and Xi Tong for the faithful discussions. T. Takahashi is supported by JSPS KAKENHI Grant Number JP23KJ1214.
T.Tanaka is supported by JSPS KAKENHI Grant Nos. JP23H00110, JP20K03928, JP24H00963, and JP24H01809.. 
H.\ O. was supported by JSPS KAKENHI Grant Number JP22J14159.
\end{acknowledgments}

\appendix

\begin{widetext}
\section{Calculation of the flux}\label{app:flux}
In this appendix, we give a general formalism to calculate the particle number fluxes of the axion at infinity and the horizon.
Here, we consider the following zeroth and first-order equations:

\begin{align}
	\left(i\partial_t+\frac{1}{2\mu}\nabla^2+\frac{\alpha}{r}\right)\psi^{(0)}&=0~, \\
	\left(i\partial_t+\frac{1}{2\mu}\nabla^2+\frac{\alpha}{r}\right)\psi^{(1)}&={\cal S}(x)~,
\end{align}
where the zeroth order solution is given by
\begin{align}
	\psi^{(0)}=\sqrt{N_0}e^{-i(\omega_0-\mu)t}R_{n_0l_0}(r)Y_{l_0m_0}(\theta,\varphi)~,
\end{align}
and ${\cal S}(x)=e^{-i\omega_s t}S(\bm{x})$ represents an oscillating source.
The specific sources used in the main text are summarized in Table.~\ref{tab:source}.

The perturbative solution can be constructed by using the Green's function, which satisfy
\begin{align}
	\left(i\partial_t+\frac{1}{2\mu}\nabla^2+\frac{\alpha}{r}\right)G(x,x')=\delta(x-x')~,
\end{align}
as
\begin{align}\label{eq:pertsol}
	\psi^{(1)}(x)=\int d^4\! x'\ G(x,x'){\cal S}(x')~.
\end{align}
We can decompose the Green's function as
\begin{align}
	G(x,x')=\sum_{lm}\int\frac{d\omega}{2\pi}e^{-i(\omega-\mu)(t-t')}Y_{lm}(\theta,\varphi)Y_{lm}^{\ast}(\theta',\varphi')G_{l\omega}(r,r')~,
\end{align}
where
\begin{align}
	G_{l\omega}(r,r')&=\frac{2\mu}{W_l(\omega)}\left(R_{kl}^{\rm in}(r)R_{kl}^{\rm up}(r')\Theta(r'-r)+R_{kl}^{\rm in}(r')R_{kl}^{\rm up}(r)\Theta(r-r')\right)~, \label{eq:Green} \\
	W_l(\omega)&=r^2\left(R_{kl}^{\rm in}\frac{dR_{kl}^{\rm up}}{dr}-R_{kl}^{\rm up}\frac{dR_{kl}^{\rm in}}{dr}\right)~.
\end{align}
Here, we impose the regularity at the origin and the outgoing boundary condition at infinity for $R_{kl}^{\rm in}(r)$ and $R_{kl}^{\rm up}(r)$, respectively. 
For the relativistic regime, the former condition is replaced with the ingoing boundary condition at the horizon.
$R_{kl}^{\rm in}$ and $R_{kl}^{\rm up}$ are the Coulomb wavefunctions, whose explicit forms are
\begin{align}
	R_{kl}^{\rm in}(r)&=C_{kl}e^{-ikr}(2ikr)^l\ _1F_1(i/kr_0+l+1,2l+2,2ikr)~, \label{eq:in_mode} \\
	R_{kl}^{\rm up}(r)&=D_{kl}e^{ikr}(-2ikr)^l\ U(-i/kr_0+l+1,2l+2,-2ikr)~, \label{eq:up_mode}
\end{align}
with
\begin{align}
	k=\sqrt{2\mu(\omega-\mu)}~, \quad r_0=\frac{1}{\mu\alpha}~,
\end{align}
where $_1F_1$ and $U$ are, respectively, the confluent hypergeometric functions of the first and second kinds~\cite{mathematicalfunction}, and $C_{kl}$ and $D_{kl}$ are normalization constants.
As a result, we obtain 
\begin{align}
    \psi^{(1)}=&\sqrt{N_0}\sum_{lm}e^{-i(\omega_1-\mu)t}\frac{2\mu}{W_l(\omega_1)}Y_{lm}(\theta,\varphi) \notag \\
    &\times\left[R_{k_1l}^{\rm up}(r)\int_0^r dr'\int d\cos\theta d\varphi\ R_{k_1l}^{\rm in}Y_{lm}^\ast S +R_{k_1l}^{\rm in}(r)\int_r^\infty dr'\int d\cos\theta d\varphi\ R_{k_1l}^{\rm up}Y_{lm}^\ast S \right]~,
\end{align}
with
\begin{align}
	\omega_1=\omega_s+\mu~, \quad k_1=\sqrt{2\mu(\omega_1-\mu)}~.
\end{align}

The particle number fluxes at infinity and the horizon are given by
\begin{align}
	{\cal F}_{{\cal I}}&=\left.\frac{r^2}{2\mu i}\int d\cos\theta d\varphi \ \left(\psi^{(1)\ast}\partial_r\psi^{(1)}-\psi^{(1)}\partial_r\psi^{(1)\ast}\right)\right|_{r\to\infty}~, \label{eq:FI} \\
	{\cal F}_{{\cal H}}&=\left.\frac{2Mr_+}{\mu}\int d\cos\theta d\varphi \left(\mu-m\Omega_H\right) \left|\psi^{(1)}\right|^2\right|_{r\to 0}~, \label{eq:FH}
\end{align}
respectively.
The asymptotic form of the perturbative solution at infinity (the horizon) is obtained from Eq.~\eqref{eq:pertsol} as
\begin{align}
	\psi^{(1)}\to\sqrt{N_0}\sum_{lm}e^{-i(\omega_1-\mu)t}\frac{2\mu}{W_l(\omega_1)}R_{k_1l}^{\rm up (in)}(r)Y_{lm}(\theta,\varphi)\int d^3\! x' R_{k_1l}^{\rm in (up)}Y_{lm}^{\ast}S~.
\end{align}
Substituting this into Eq.~\eqref{eq:FI} and \eqref{eq:FH}, we obtain the resultant expression of the fluxes as
\begin{align}
	{\cal F}_{{\cal I}}&=4\mu k_1 \sum_{lm}\frac{1}{\left|W_l(\omega_1)\right|^2}\left.r^2|R_{k_1l}^{\rm up}|^2\right|_{r\to\infty} \left|\int d^3\! x' R_{k_1l}^{\rm in}Y_{lm}^{\ast}S\right|^2 N_0~, \label{eq:FI2} \\
	{\cal F}_{{\cal H}}&=8Mr_+\mu \sum_{lm}\frac{\mu-m\Omega_H}{\left|W_l(\omega_1)\right|^2}\left|R_{k_1l}^{\rm in}(0)\right|^2\left|\int d^3\! x' R_{k_1l}^{\rm up}Y_{lm}^{\ast}S\right|^2 N_0~. \label{eq:FH2}
\end{align}

\begin{table*}\label{tab:source}
\caption{\label{tab:source}Source terms.}
\begin{tabular}{|c|c|}
\hline
 Self-interaction & \begin{tabular}{c} ${\cal S}=-\frac{1}{8F_a^2}\psi_{211}^\ast\psi_{322}^2$ (Infinity) \\ ${\cal S}=-\frac{1}{8F_a^2}\psi_{211}^2\psi_{322}^\ast$ (Horizon) \end{tabular} \\ \hline
 Tidal interaction & ${\cal S}=V_{l_{\ast},m_{\ast}}e^{\mp i m_{\ast}\Omega t}\psi_i \quad (i=1,2)$\\ \hline
\end{tabular}
\end{table*}

\section{Perturbation theory v.s. two-mode approximation}\label{app:two}
In this appendix, we compare the perturbation theory and the two-mode approximation regarding the calculation method of dissipation fluxes, especially for the indirect emission induced by the tidal interaction mentioned in Sec.~\ref{subsubsec:indirect}.
As we formulated in Ref.~\cite{Takahashi:2023flk}, we evaluate the indirect decay rate by considering a limited configuration composed of only two modes, 
\begin{align}
    \psi=c_i(t)\psi_i+c_d(t)\psi_d~,
\end{align}
where $c_{i,d}$ are complex coefficients, $\psi_i$ and $\psi_d$ represent the initial and intermediate modes, respectively. 
We consider the situation in which the intermediate mode is unstable with the decay rate $f\equiv{\cal F}_{d,{\cal I}}^{\rm td}/2N_d$. 
From the equation of motion with the tidal potential, the evolution equation for the coefficients can be rewritten as
\begin{align}\label{eq:two_eom}
    i\frac{d}{dt}\begin{pmatrix}c_i \\ c_d \end{pmatrix}=\begin{pmatrix}
        \delta\Omega+i\omega_{i,I} & \eta \\
        \eta & -\delta\Omega-if
    \end{pmatrix}\begin{pmatrix}c_i \\ c_d \end{pmatrix}
\end{align}
where
\begin{align}
    \delta\Omega&=\pm\frac{m_{\ast}}{2}(\Omega-\Omega_{\rm res})~, \\
    \eta&=\left|\int d^3x R_{n_dl_d}Y_{l_d,m_d}^{\ast}V_{l_{\ast}m_{\ast}}R_{n_il_i}Y_{l_i,m_i}\right|~,
\end{align}
with $\Omega_{\rm res}=\pm((\omega_R)_d-(\omega_{R})_i)/(m_d-m_i)$. 
The eigenvalues of the matrix on the right hand side are
\begin{align}
    \lambda_{\pm}=&\frac{i}{2}(\omega_{i,I}-f)\pm\frac{1}{2}\sqrt{\{2\delta\Omega+i(\omega_{i,I}+f)\}^2+4\eta^2}~.
\end{align}
If the transition is adiabatic, the eigenstate of axions originally at $\psi_i$ mode follows the one with the eigenvalue $\lambda_+$. 
\footnote{Expanding the second term of $\lambda_+$ to the first order with respect to $\eta/|2\delta\Omega+i(f+\omega_{i,I})|$ and taking the imaginary part, one can see that the coefficient of the second term of the right-hand side of Eq.~(39) in Ref.~~\cite{Takahashi:2023flk} is reproduced.}
Therefore, the number flux of the indirect emission is estimated as 
\begin{align}
    {\cal F}_{i,{\rm Ind}}^{\rm td}=2(\Im{\lambda_+}-\omega_{i,I})N_i~.
\end{align}

We should note that the two-mode approximation is justified only around the resonance between the two modes.
On the other hand, in the main text, we evaluate this process using the second-order perturbation theory.
In Fig.~\ref{fig:PertTwo}, we show the flux calculated by each method.
Here, we ignore $\omega_{i,I}$ when calculating the flux for simplicity.
There are two qualitative differences between the results of these two methods.
First, the result of the two-mode approximation has a finite value at a lower frequency even when the flux evaluated by the second-order perturbation theory vanishes.
In the former approach, the flux has a finite value if the decay rate of the destination excited bound mode does not vanish.
Thus, the threshold orbital frequency is $(\mu-\omega_{1,R})/m_\ast$.
On the other hand, in the latter second-order perturbation, to excite unbound modes, the orbital frequency must be larger than $(\mu-\omega_{1,R})/(m_\ast^{(1)}+m_\ast^{(2)})$, where $m_\ast^{(1)}$ and $m_\ast^{(2)}$ are the magnetic quantum number of the tidal potential involved in the first and second transitions, respectively.

Second, the behavior around the resonance is different.
Near the resonance, the perturbation theory breaks down, because the particle number of the first-order solution $\psi^{(1)}$, which represents the intermediate state of this two-step transition, formally diverges. 
This divergence is caused by crossing zero of the Wronskian at the resonance.
In Fig.~\ref{fig:N1}, we show the particle number $N^{(1)}=\int d^3x|\psi^{(1)}|^2$ as a function of the orbital frequency.
The perturbative expansion is unreliable in the region where the $N^{(1)}$ exceeds the original particle number $N_0$.
Instead, around the resonance, the two-mode approximation should work well.
Fig.~\ref{fig:PertTwo} shows that when the perturbation theory is valid over a wide range, as in the case of $q=10^{-2}$, the two results are almost identical around the resonances. 

To summarize, the two-mode approximation is useful when we study the evolution around the resonance.
However, when applied to low-frequency region away from the resonance, it unphysically overestimates the flux.
In such a region we should rely on the second-order perturbation theory.

\begin{figure*}[t]
\includegraphics[keepaspectratio, scale=0.85]{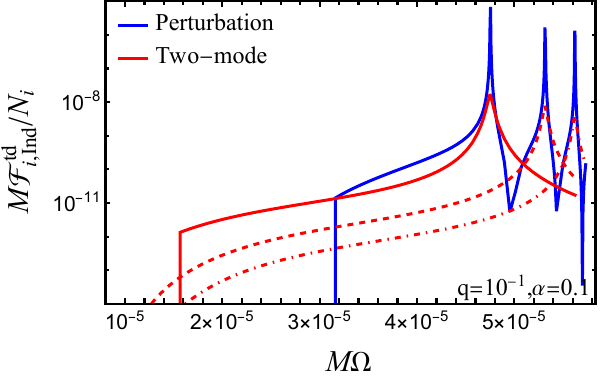}
\includegraphics[keepaspectratio, scale=0.85]{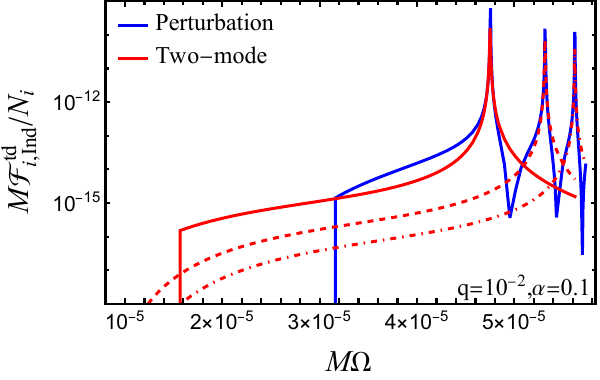}
\caption{Comparison of the fluxes of the indirect emission calculated by using the second-order perturbation theory and the two-mode approximation for $\alpha=0.1$, $q=10^{-1}$ (left) and $q=10^{-2}$ (right). The original mode is $(n,l,m)=(2,1,1)$ and we consider the excitation by a tidal potential of $l_\ast=m_\ast=2$. The blue curve shows the flux calculated by the second-order perturbation method mentioned in Sec.~\ref{subsubsec:indirect}. The intermediate mode is $l=m=3$ and the final unbound mode is $l=m=5$. The red curves show the fluxes calculated by the two-mode approximation. The solid, dashed and dash-dotted lines show the results when the transition destination mode is $(n_d,l_d,m_d)=(4,3,3), (5,3,3)$ and $(6,3,3)$, respectively.}
\label{fig:PertTwo}
\end{figure*}

\begin{figure}
\includegraphics[scale=1]{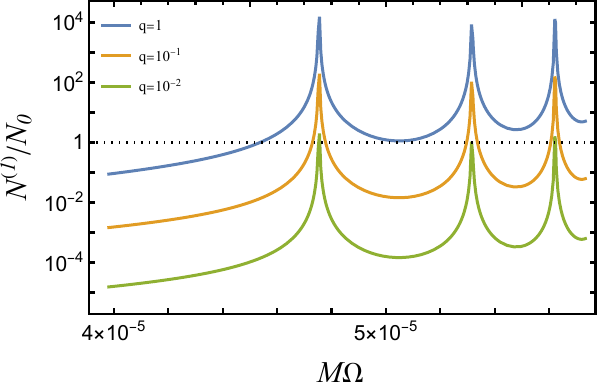}
\caption{Particle number of the first order solution in the perturbation theory for $\alpha=0.1$ with various $q$.}
\label{fig:N1}
\end{figure}

\section{Criterion for dynamical instability}\label{app:BN}
In this appendix, we summarize the toy model that explains the dynamical instability of the axion cloud originally presented in Ref.~\cite{Yoshino:2012kn} and simplified in Ref.~\cite{Omiya:2022gwu}.
Here, we consider only the self-interaction given by Eq.~\eqref{eq:cos} and the gravity from the central BH.
The potential of the axion cloud is given by
\begin{align}\label{eqapp:pot}
    V=\int d^3\! x\left(\frac{1}{2\mu}|\partial_i\psi|^2-\frac{\alpha}{r}|\psi|^2-F_a^2\mu^2\sum_{n=2}^{\infty}\frac{(-1/2)^n}{(n!)^2}\frac{|\psi|^{2n}}{F_a^{2n}\mu^n}\right)~.
\end{align}
For simplicity, we take the ansatz 
\begin{align}
    \psi(r,\theta,\varphi;A_p,r_p,\sigma)=A_p\exp\left(-\frac{(r-r_p)^2}{4\sigma^2}\right)Y_{11}(\theta,\varphi)~,
\end{align}
where the configuration of the cloud is characterized by the peak amplitude $A_p$, the position of the peak $r_p$, and the radial extension of the cloud $\sigma$.
The particle number is given by
\begin{align}
    N=\int d^3\! x|\psi|^2\simeq \sqrt{2\pi}\sigma (r_p^2+\sigma^2)A_p^2~.
\end{align}
Substituting it to Eq.~\eqref{eqapp:pot}, we obtain
\begin{align}
    \frac{V(r_p,\sigma,N)}{N}=&\frac{r_p^2+3\sigma^2}{8\mu\sigma^2(r_p^2+\sigma^2)}+\frac{1}{\mu(r_p^2+\sigma^2)}-\frac{\alpha r_p}{r_p^2+\sigma^2} \notag \\
    &-\mu^2\left(\frac{N_{\ast}}{160\pi\sqrt{2\pi}\mu^4\sigma(r_p^2+\sigma^2)}-\frac{3N_{\ast}^2}{17920\pi^3\mu^7\sigma^2(r_p^2+\sigma^2)^2}+\cdots\right)~,
\end{align}
where we defined $N_{\ast}=(\mu^2/F_a^2)N$.
When the particle number $N$ is given, the configuration of the cloud is determined by
\begin{align}
    \partial_{r_p}V=\partial_{\sigma}V=0~.
\end{align}
Eliminating $N_{\ast}$, we can solve them for $\sigma$, and denote the solution as $\sigma_{\rm eq}(r_p,N)$.
Also, we denote the potential substituted it as $V_{\rm eq}(r_p,N)=V(r_p,\sigma_{\rm eq},N)$.
For fixed $N$, the position of the peak amplitude is determined by $\partial_{r_p}V_{\rm eq}=0$.
As the particle number $N$ increases from a sufficiently small value, the position $r_p$ decreases.
However, for a certain range of $N$, three solutions of $r_p$ appear, two of which are stable points with the gravity of the BH and attractive self-interaction.
The point at which the original stable point disappears as $N$ increases can be regarded as the criterion for the dynamical instability.
Such point satisfy
\begin{align}
    \left.\frac{\partial N}{\partial r_p}\right|_{\frac{\partial V_{\rm eq}}{\partial r_p}=0}=-\left.\frac{\partial^2 V_{\rm eq}}{\partial r_p^2}\middle/\frac{\partial^2 V_{\rm eq}}{\partial r_p\partial N}\right.=0~.
\end{align}
Therefore, the solution with the larger $N$ in the following equations:
\begin{align}
    \frac{\partial V_{\rm eq}}{\partial r_p}=\frac{\partial^2 V_{\rm eq}}{\partial r_p^2}=0~,
\end{align}
gives the criterion. 
The numerical result is shown in Fig~.\ref{fig:BNth}, and it can be approximated as
\begin{align}\label{eq:BNth}
    N_{\rm BN}\simeq 246\ \alpha^{-1}\left(\frac{F_a}{\mu}\right)^2~.
\end{align}

\begin{figure}
\includegraphics[scale=1]{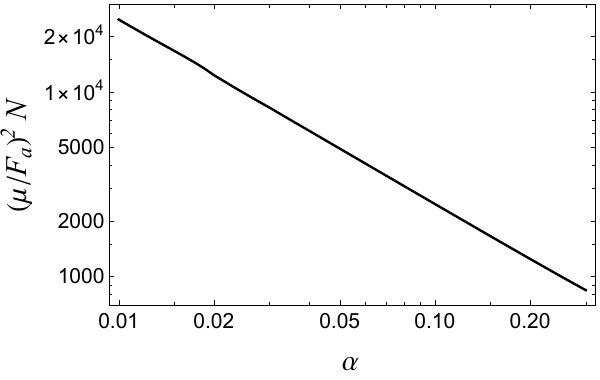}
\caption{Threshold value of the particle number at which the dynamical instability, or bosenova, occurs.}
\label{fig:BNth}
\end{figure}

\end{widetext}

\bibliography{ref}

\end{document}